\documentclass[sn-mathphys]{sn-jnl}

\usepackage[subnum]{cases}
\usepackage{multirow}

\jyear{2022}%

\theoremstyle{thmstyleone}%
\newtheorem{theorem}{Theorem}
\newtheorem{proposition}[theorem]{Proposition}%

\theoremstyle{thmstyletwo}%
\newtheorem{example}{Example}%
\newtheorem{remark}{Remark}%

\theoremstyle{thmstylethree}%
\newtheorem{definition}{Definition}%

\newcommand{\tr}{ \text{tr} }

\newcommand{\erf}{ \text{erf} }

\newcommand{\Ima}{ \text{Im} }
\newcommand{\cl}{ \text{cl} }
\newcommand{\qm}{ \text{q} }

\newcommand{\ti}{ \tilde }

\newcommand{\ep}{ \epsilon }
\newcommand{\pa}{ \partial }
\newcommand{\hb}{ \hbar }
\newcommand{\si}{ \sigma }
\newcommand{\ga}{ \gamma }
\newcommand{\om}{ \omega }

\newcommand{\del}{ \delta }
\newcommand{\la}{ \langle }
\newcommand{\ra}{ \rangle }

\begin{document}

\title[Stochastic Bohmian and Scaled Trajectories]{Stochastic Bohmian and Scaled Trajectories}


\author[1]{\fnm{S. V.} \sur{Mousavi}}\email{vmousavi@qom.ac.ir}
\equalcont{These authors contributed equally to this work.}

\author*[2]{\fnm{S.} \sur{Miret-Art\'es}}\email{s.miret@iff.csic.es}
\equalcont{These authors contributed equally to this work.}


\affil[1]{\orgdiv{Department of Physics}, \orgname{University of Qom}, \orgaddress{\street{Ghadir Blvd.}, \city{Qom}, \postcode{371614-6611}, \country{Iran}}}

\affil[2]{\orgdiv{Instituto de F\'isica Fundamental}, \orgname{Consejo Sperior de Investigaciones Cient\'ificas}, \orgaddress{\street{Serrano 123}, \city{Madrid}, \postcode{28006}, \country{Spain}}}



\abstract{In this review we deal with open (dissipative and stochastic) quantum systems within the Bohmian mechanics framework which has the 
advantage to provide a clear picture of quantum phenomena in terms of trajectories, originally in configuration space.  The gradual 
decoherence process is studied from linear and nonlinear Schr\"odinger equations through Bohmian trajectories as well as by using the so-called quantum-classical  transition differential  equation through scaled trajectories. This 
transition is governed by a continuous parameter, the transition parameter, covering these two extreme open dynamical regimes. 
Thus, two sources of decoherence of different nature are going to be considered. Several examples will 
be presented and discussed in order to illustrate the corresponding theory behind each case, namely: the so-called Brownian-Bohmian 
motion leading to quantum diffusion coefficients, dissipative diffraction in time,
dissipative tunnelling for a parabolic barrier under the presence of an electric field and stochastic early arrivals for the same type of barrier. 
In order to simplify the notations and physical discussion, the theoretical developments will be carried out in one dimension throughout all 
this wok. One  of the main goals is to analyze the gradual decoherence process existing in these open dynamical regimes in terms of 
trajectories, leading to a more intuitive way of understanding the  underlying physics in order to gain new insights. }





\maketitle

\section{Introduction}\label{sec1}

Real physical systems do not exist in complete isolation in nature and therefore they can be thought as  open systems 
in the sense that the interaction with their environments is always present. From the very beginning, the motion of particles in quantum 
mechanics was analyzed in terms of  stochastic processes. A formal analogy between the Brownian motion and the Schr\"odinger equation was 
noticed by F\"urth \cite{furth}. Subsequently, F\'enyes \cite{fenyes} and Weizel \cite{weizel1,weizel2,weizel3}
developed  this approach with more mathematical detail. The search for a stochastic support for quantum  mechanics comes from the early 
1950s. The first attempt in the Bohmian mechanics was due to Bohm and Vigier \cite{bohm3}. These authors assumed that the
electron is a particle suspended in a Madelung fluid \cite{madelung} whose general motion is determined by the resolution of the 
Schr\"odinger equation.  In general, the aim was to show how close it is to 
classical theory of Brownian motion and Newtonian mechanics, and how the Schr\"odinger 
equation might have been discovered from this point of view. The theory was well established by
Nelson \cite{Ne-PR-1966} starting from a different point of view of those proposed by 
F\'enyes, Weizel, Kershaw \cite{kershaw}, Comisar \cite{comisar} and de la Pe\~na
\cite{pena}. Olavo \cite{olavo} also followed previous works but starting from a more axiomatic formulation. 
In this respect, one can claim that the depart of Bohm and collaborators is again a breakthrough. Bohm and Hiley \cite{bohm4,bohm5}  
make clear that there is no way to avoid non-locality in  the stochastic interpretation of 
quantum mechanics. Recently, this initial work has also been reviewed and extended by Drezet within the context of Brownian motion in the pilot wave interpretation of de Broglie  \cite{drezet}.

In general, the field of open quantum systems is growing very rapidly due to the fact that the interaction of the system with an environment and 
the entanglement between both lead to the emergence of a new and very exciting physics \cite{Percival1998,Weiss-book-1999,BaPe-book-2002,Joos-2003,Sch-2007,NaMi-book-2017}.
There are three main approaches in the literature to deal with the dissipative and stochastic dynamics in open quantum systems \cite{Weiss-book-1999,NaMi-book-2017}:
(i) the system-plus-environment approach leading to master equations for the reduced density matrix which is issued from integration
over the environmental degrees of freedom, (ii) effective time-dependent Hamiltonians through linear Schr\"odinger equations such as the so-called Caldirola-Kanai (CK) equation \cite{Caldirola, Kanai} and
(iii) non-linear Schr\"{o}dinger equations. We are going to focus here on the second and third approaches where the wave function is  
the object of interest, not the reduced density matrix. Seven of such non-linear Schr\"{o}dinger equations have been analyzed elsewhere \cite{BaNa-JAMA-2012} 
by providing the corresponding Feynman propagators: the Bialynicki-Birula and Mycielski (BBM) equation \cite{BBMy-AP-1976}, 
the Bateman-Caldirola-Kanai equation \cite{Bateman,Caldirola,Kanai}, the Di\'osi-Halliwell-Nassar equation \cite{Diosi,BaNa-JAMA-2012}, 
the Schr\"{o}dinger-Langevin (SL) or Kostin equation \cite{Ko-JCP-1972}, the Schuch-Chung-Hartmann (SCH) equation \cite{ScChHa-JMP-1983, Sc-IJQC-1999}, 
the S\"ussmann-Hasse-Albretch-Kostin-Nassar equation \cite{Hasse,Albrecht,Kostin,Nassar1} and  
the Schr\"odinger-Nassar equation \cite{Nassar2}.
Among these nonlinear  Schr\"odinger equations, the SL equation derived heuristically by Kostin  from the Heisenberg-Langevin equation for the 
momentum operator seems to be the most widely used. 
Furthermore, a nonlinear Schr\"{o}dinger equation was also proposed by Nassar and Miret-Art\'es (NMA) \cite{NaMi-PRL-2013} to describe 
at the same time the continuous measurement of the position of a quantum particle interacting with its environment. 
Chavanis \cite{Ch-EPJP-2017} also derived two other nonlinear equations using the theory of scale relativity due to Nottale \cite{No-book-2011}
by introducing a complex friction coefficient. For real frictions, one of them reduces to the SL equation.
%
A different generalized SL
equation was proposed in the literature for nonlinear interaction providing a state-dependent dissipation process exhibiting multiplicative noise
\cite{BaMi-AOP-2014}. This differential equation was afterwards extended to a non-Markovian scheme \cite{VaMoBa-AOP-2015}.

On the other hand, Bohmian mechanics \cite{Holland-book-1993,salva1,salva2} is clearly a complementary,  alternative and new interpretive way of introducing quantum mechanics and has the 
advantage to provide a clear picture of quantum phenomena in terms of trajectories, originally in configuration space.  
Our purpose is also to present and analyze the corresponding trajectories in order to describe open dynamics within this formalism 
(we will speak about dissipative and stochastic Bohmian trajectories). Furthermore, if the gradual decoherence process is also studied 
by using the so-called quantum-classical 
transition wave differential equation, originally proposed by Richardson et al. \cite{RiSchMaVaBa-PRA-2014} 
in the context of conservative systems, the corresponding dynamics are governed by a continuous parameter, the transition parameter. 
Chou  applied this wave equation to analyze wave-packet  interference \cite{Chou1-2016}  and the dynamics of the harmonic 
and Morse oscillators with complex trajectories \cite{Chou2-2016}.
The resulting trajectories have been called by us {\it scaled trajectories} 
\cite{MoMi-AP-2018,MoMi-JPC-2018} since a scaled Planck's constant in terms of the transition parameter is used. 
Moreover, by assuming a time-dependent Gaussian ansatz for the probability density,
Bohmian and scaled trajectories are expressed as a sum of  a classical trajectory (a particle property) 
and a term containing the width of the corresponding wave packet (a wave property) within 
of what has been called the {\it dressing scheme} \cite{NaMi-book-2017}. Similarly, the same scheme is observed in the context of 
nonlinear quantum mechanics \cite{Feng} where solitons, which differ completely 
from normal microscopic particles, also possess the wave-particle duality; their wave property appears in the form of a travelling solitary 
wave and their corpuscle feature is analogous to a classical particle. 
Scaled trajectories also display the well-known non-crossing property even in the classical regime. 
%

The procedure of using a continuous parameter monitoring the different dynamical regimes in the theory could be seen quite similar 
to the WKB approximation (based on a series expansion in powers of Planck's constant), widely used for conservative systems. However, 
some important differences should be clearly stressed.
First, whereas the classical Hamilton-Jacobi equation for the classical action is obtained at zero order in the WKB approximation,  
the so-called classical wave (nonlinear) equation \cite{Schi-PR-1962} is reached by construction. 
Second, the hierarchy of the differential equations for the action at different orders of the expansion in $\hbar$ is substituted by only a
transition wave differential equation which can be  solved in the linear and nonlinear domains. Third, 
the transition from quantum to classical trajectories is carried out in a continuous and gradual way. Fourth, the scaling procedure 
extended and applied to open quantum systems is very easy to implement.
And fifth, the gradual decoherence process due to the scaled Planck's constant allows us to analyze what happens at intermediate regimes.
At this level, it is interesting to point out that this process studied within this theoretical scheme displays two sources of decoherence but of 
different nature; the first one is due to the open dynamics itself and the second one due to the transition parameter pinpointing and 
stressing different dynamical regimes until reaching the open classical regime.

In the nonlinear Schr\"odinger equation approach,  the so-called  Schr\"{o}dinger-Langevin-Bohm (SLB) equation is obtained by 
substituting the polar form of the wave function into the SL nonlinear differential equation \cite{NaMi-book-2017}. Then, by assuming a 
Gaussian ansatz as a solution of the SLB equation  for potentials of at most second order in space coordinate, the center of the wave packet 
is ruled by the {\it classical} Langevin equation and its width fulfills the so-called generalized Pinney equation \cite{Pinney}. Surprisingly,
thermal fluctuations do not affect the width of the wave packet. 
The classical Langevin equation is solved by using a white noise for the stochastic force which is typical for the 
Brownian motion. In Ref.  \cite{KaGo-AOP-2016}, quantum noise (white and colored) has been used to solve the Kostin equation for 
the simple harmonic oscillator. 
The most basic problem to be considered is the Brownian motion described by stochastic Bohmian trajectories. Thus, one speaks 
about the Brownian-Bohmian motion. In this diffusion regime, time dependent classical and quantum diffusion coefficients issued from 
mean square displacements in the classical and Bohmian context are defined, respectively. At asymptotic times, both diffusion coefficients 
tend to be the same. This behavior can also be derived from the velocity autocorrelation function \cite{Mi-2018}. 

In this work, several examples will be presented and discussed in order to illustrate the corresponding theory behind each case, 
namely: the Brownian-Bohmian motion leading to quantum diffusion coefficients, dissipative diffraction in time reported when a sudden 
release from a totally absorbing shutter occurs in a free non-conservative dynamics (originally it was reported for conservative dynamics),
dissipative tunnelling for a parabolic barrier under the presence of an electric field and stochastic early arrivals with respect 
to free propagation for the same type of  barriers but time dependent. In particular,  
dissipative tunnelling through a parabolic repeller has been studied via quantum Langevin equation \cite{FoLeCo-PLA-1988}, 
in resistively shunted Josephson junctions, and using the CK \cite{Pa-JPA-1997} and Kostin \cite{MoMi-AP-2018} approaches.
Investigation of dissipative tunnelling times through different approaches have attracted many researchers 
\cite{KoNiTaHa-PLA-2007, BhRo-PRA-2012, BhRo-JMP-2013, KeLoEd-AOP-2017, Po-PRL-2017}.  Thus, Baskoutas and Jannussis 
\cite{BaJa-JPA-1992} and Papadopoulos \cite{Pa-JPA-1997}, Tokieda and Hagino \cite{ToHa-PRC-2017} have considered  
similar theoretical frameworks.  In order to simplify the notations and physical discussions, the theoretical
developments will be carried out here in one dimension. 
One  of the main goals is to illustrate the gradual decoherence process existing in these  dynamical regimes  by
applying very different theoretical methods to those widely used such as the density matrix, path-integral and 
Langevin formalisms \cite{Weiss-book-1999}.  In this different and less known alternative  \cite{NaMi-book-2017}, the decoherence
process is followed by means of stochastic trajectories leading to a more intutitive way of understanding the underlying physics. In this sense, 
analytically solvable models are very useful in order to gain new insights.

\section{Dissipative and Stochastic Bohmian trajectories}\label{sec2}

\subsection{The Caldirola-Kanai equation}\label{subsec2-1}

In this approach, the Lagrangian  for a one particle system is given by \cite{Razavy-2005}
\begin{eqnarray*} \label{eq: class_lagrange}
	\mathcal{L} &=& \left[ \frac{m}{2} \dot{x}^2 - V(x)  \right] e^{\gamma t} 
\end{eqnarray*}
where $m$ is the mass of the particle, $V$ the external potential and $ \ga $ the friction coefficient. Here we follow the standard notation 
where a dot on a given variable means derivation with respect to time. The  canonical momentum  $p_c$ is obtained as
\begin{eqnarray} \label{eq: class_canonical momentum}
p_c &=& \frac{\partial \mathcal{L}}{\partial \dot{x}} = m \dot{x} e^{\gamma t} ,
\end{eqnarray}
which is explicitly time-dependent and $m \dot{x} = p_c e^{-\gamma t}$ is the kinematic momentum.
The corresponding Hamiltonian function is given by 
\begin{equation}
H_{CK} = \dot{x} p_c - L = e^{-\gamma t} \frac{p_c^2}{2m} + e^{\gamma t} V(x)  .
\end{equation}
The CK Hamiltonian operator $\hat{H}_{CK}$ can then be obtained from the standard quantization 
rule by substituting the canonical momentum $p_c$ by $\frac{\hbar}{i} \frac{\partial}{\partial x}$, that is,
\begin{eqnarray} \label{eq: KC QM-Hamiltonian}
	\hat{H}_{CK} &=& - \frac{\hbar^2}{2m} e^{-\gamma t} \frac{\partial^2}{\partial x^2} + 
	e^{\gamma t} V(x) .
\end{eqnarray}
The commutation relation is $[x, p_c] = i \hbar$ and the uncertainty principle is formally fulfilled, $\Delta x \Delta p_c \sim \hbar $.
Notice however that  the commutation relation  of the position operator and kinematic momentum is
$i \hbar e^{-\gamma t}$.
Thus,  as long as quantities related to this momentum are not
computed, the use of this wave differential equation in the physical coordinate space is formally correct
\cite{SaCaRoLoMi-AP-2014}. 
%
The time-dependent Schr\"{o}dinger equation within this CK approach then reads  as
\begin{eqnarray} \label{eq: Sch_viscid}
	i \hbar \frac{\partial}{\partial t}\psi(x, t) &=& \left[ -\frac{\hbar^2}{2m} e^{-\gamma t}
	\frac{\partial^2}{\partial x^2} + V(x) e^{\gamma t}
	\right] \psi(x, t) .
\end{eqnarray}
Now, if the polar form of the wave function is used as in standard Bohmian mechanics  and written as
\begin{equation}\label{Psi}
\psi(x, t) = R(x, t) e^{iS(x, t)/\hbar}  ,
\end{equation}
Eq. (\ref{eq: Sch_viscid}) is then split in two coupled differential equations by separating the real and imaginary parts according to 
\begin{eqnarray}
	\frac{\partial  R^2}{\partial t} + \frac{\partial}{\partial x}\left( R^2 \frac{1}{m} 
	\frac{\partial  S}{\partial x} e^{-\gamma t} \right)  &=& 0 , \label{eq: continuity_viscid} \\
	\frac{\partial  S}{\partial t} + \frac{1}{2m} \left( \frac{\partial  S}{\partial x} \right)^2 
	e^{-\gamma t} + ( V(x) + Q e^{- 2\gamma t} ) e^{\gamma t} &=& 0 , \label{eq: HamJac_viscid}
\end{eqnarray}
with $ Q(x, t) $ defined as
\begin{eqnarray} \label{eq: Qp_viscid}
	Q(x, t) &=&  -\frac{\hbar^2}{2m} \frac{1}{R}\frac{\partial^2 R}{\partial x^2} , 
\end{eqnarray}
$ Q(x, t) e^{-2 \gamma t} $ being the quantum potential in this CK framework.
Eqs. (\ref{eq: continuity_viscid})  and (\ref{eq: HamJac_viscid}) are the well-known continuity  and  Hamilton-Jacobi equations for dissipative 
problems respectively, and the probability density and current density are expressed as 
\begin{eqnarray}
	\rho (x, t) &=& R^2(x, t)  , \label{eq: den_prob} \\
	j(x, t) &=& R^2 \frac{1}{m} \frac{\partial  S}{\partial x} e^{-\gamma t}  =
	\frac{\hbar}{m}~ \Ima \left\{ \psi^* \frac{\partial \psi}{\partial x} \right\} e^{-\gamma t}  ,
	 \label{eq: cur_den_prob}
\end{eqnarray}
where $\Ima$ stands for the imaginary part.
Furthermore, the dissipative Bohmian velocity field is given by the guiding condition 
\begin{eqnarray} \label{eq: velocity_field}
	v_B(x, t) &=& \frac{j(x, t)}{\rho(x, t)} = \frac{1}{m} \frac{\partial  S}{\partial x}  e^{-\gamma t} ,
\end{eqnarray}
from which dissipative Bohmian trajectories $x(x^{(0)}, t)$ are extracted by integrating this equation leading to 
\begin{eqnarray} \label{eq: guid_viscid}
	\frac{dx}{dt} &=& v_B(x, t) \bigg\vert_{x=x(x^{(0)}, t)} .
\end{eqnarray}
As can be seen, the only initial condition is the 
position of the particle since its initial momentum is specified by the phase of the wave 
function according to Eq. (\ref{eq: velocity_field}). From Eq. (\ref{eq: guid_viscid}), one has that 
\begin{eqnarray} \label{eq: acceleration_viscid}
	m\ddot{x} &=& \frac{d}{dt} \left(  e^{-\gamma t} \frac{\partial  S}{\partial x}  \right) =
	e^{-\gamma t} \left[ -\gamma \frac{\partial S }{\partial x} +
	\left( \frac{\partial  }{\partial t} + \dot{x} \frac{\partial  }{\partial x} \right) \frac{\partial S }{\partial x}
	\right] .
\end{eqnarray}
Now, from Eqs. (\ref{eq: guid_viscid}) and (\ref{eq: HamJac_viscid}),  
Eq. (\ref{eq: acceleration_viscid}) can then be rewritten as
\begin{eqnarray} \label{eq: acceleration_viscid2}
	m\ddot{x} + m \gamma \dot{x} + \frac{\partial}{\partial x} (V + e^{-2\gamma t} Q) &=& 0 ,
\end{eqnarray}
which resembles the classical Langeving equation without noise except for the extra term 
$e^{-2\gamma t} \partial Q / \partial x$ which represents the quantum force derived from the
quantum potential, Eq. (\ref{eq: Qp_viscid}). 

\subsection{The Schr\"odinger-Langevin or Kostin equation}\label{subsec2-2}

Kostin derived heuristically the so-called Schr\"{o}dinger-Langevin  equation which  reads
as \cite{Ko-JCP-1972,Kostin}
\begin{eqnarray} \label{eq: Sch_Lan}
	i \hb \frac{\pa}{\pa t}\psi &=& \left[ -\frac{\hb^2}{2m} \frac{\pa^2}{\pa x^2}
	+ V + V_r + \frac{\gamma \hb }{2 i} \left( \ln \frac{\psi}{\psi^*} - \left \langle \ln \frac{\psi}{\psi^*} \right \rangle \right) 
	\right] \psi   ,
\end{eqnarray}
which is widely used 
to describe open quantum systems using the quantum Langevin equation. Here,  $V(x,t)$ and $V_r(x,t)$ are the external and  
random potential depending also on time, respectively. Averaged values calculated from the wave function are represented by $\langle \cdots \rangle$.
The random potential is assumed to be linear on the particle position  according to
\begin{eqnarray} \label{eq: random_pot}
	V_r(x, t) &=& x~F_r(t) ,
\end{eqnarray}
$F_r(t)$ being a time-dependent random force. 
By substituting again Eq.(\ref{Psi})
%
%
into the SL or Kostin equation (\ref{eq: Sch_Lan}), the resulting Schr\"{o}dinger-Langevin-Bohm (SLB) equation is written 
as \cite{NaMi-book-2017,Mousavi-Salva-2019}
\begin{eqnarray} \label{kostin-bohmian} 
	- \frac{\pa S}{\pa t} + i \hb \frac{1}{R} \frac{\pa R}{\pa t} 
	& = & 
	\frac{1}{2m} \left( \frac{\pa  S}{\pa x} \right)^2 + 
	V + V_r + \gamma ( S - \langle S \rangle ) - \frac{\hb^2}{2m} \frac{1}{R} \frac{\pa^2 R}{\pa x^2}
	\nonumber \\
	&-& i \hb \frac{1}{2m} \left( \frac{\pa^2 S}{\pa x^2} + \frac{2}{R} \frac{\pa R}{\pa x} \frac{\pa S}{\pa x} \right)   ,
\end{eqnarray}
and following the same procedure as before, the corresponding coupled equations are 
\begin{eqnarray} \label{eq: con}
	\frac{\pa  \rho}{\pa t} + \frac{ \pa }{\pa x} \left( \rho \frac{1}{m} \frac{\pa S}{\pa x} \right) &=&  0 , 
\end{eqnarray}
and
\begin{eqnarray}  \label{eq: HJ}
	- \frac{\pa  S}{\pa t} &=& \frac{1}{2m} \left( \frac{\pa  S}{\pa x} \right)^2 + 
	V + V_r + \gamma ( S - \langle S \rangle ) + Q  ,
\end{eqnarray}
which are again the continuity and Hamilton-Jacobi equations, respectively.
%
From the velocity field or guiding equation
\begin{eqnarray} \label{eq: BM_vel}
	v(x, t) &=& \frac{1}{m} \frac{\pa S}{\pa x} 
\end{eqnarray}
and the Hamilton-Jacobi equation, a Langevin-type equation is then easily deduced
\begin{eqnarray} \label{eq: Newton}
	\frac{d v}{dt} &=& - \ga v - \frac{1}{m} \frac{\pa}{\pa x}( V + Q ) - \frac{F_r}{m}   ,
\end{eqnarray}
where the relation
\begin{eqnarray} 
	\frac{d}{dt} &=& \frac{\pa}{\pa t} + v \frac{\pa}{\pa x}
\end{eqnarray}
has been used. Stochastic Bohmian trajectories are derived from this Langevin-type equation due to the presence of dissipation and noise.

When the random force is neglected, the resulting differential equation can be seen as an alternative description of the dissipative 
dynamics compared to the CK view. The differences between both types of trajectories will be illustrated in Section \ref{sec5}.

\subsection{The Generalized nonlinear Schr\"odinger equation}\label{subsec2-3}

Based on  the theory of scale relativity due to Nottale, Chavanis \cite{Ch-EPJP-2017, No-book-2011} derived the following quantum equation
of motion
\begin{eqnarray} \label{eq: der_HJ}
	\frac{\partial}{\partial x} \left \{ \frac{ \pa \mathcal{S} }{ \pa t } + \frac{1}{2m} \left( \frac{\partial \mathcal{S}}{\partial x} \right)^2 - i \mathcal{D} \frac{\partial^2 \mathcal{S}}{\partial x^2}  + V + V_r + \ga \mathcal{S} \right\} &=& 0  ,
\end{eqnarray}
with $\mathcal{S}$ being a complex action, $\mathcal{D}$ playing the role of a diffusion coefficient and $\ga$ a complex friction.
After integrating over space, the Hamilton-Jacobi equation within this framework is written as
\begin{eqnarray} \label{eq: HJ}
	\frac{ \pa \mathcal{S} }{ \pa t } + \frac{1}{2m} \left(\frac{\partial \mathcal{S}}{\partial x} \right)^2 - i \mathcal{D} \frac{\partial^2 \mathcal{S}}{\partial x^2} + V + V_r + \ga \mathcal{S} + f(t) &=& 0   ,
\end{eqnarray}
where $f(t)$ is a constant of integration which will be determined later on. By introducing now the wave function as follows 
(the so-called Schr\"odinger transformation)
\begin{eqnarray} \label{eq: wf}
	\mathcal{S} &=& -2 i m \mathcal{D} \ln \psi    ,
\end{eqnarray}
Eq. (\ref{eq: HJ}) can be rewritten as
\begin{eqnarray} \label{eq: eq1}
	2 i m \mathcal{D} \frac{\pa \psi}{\pa t} &=& \left\{ -2 m \mathcal{D}^2 \frac{\pa^2}{\pa x^2}  + V + V_r - 2 i m \mathcal{D} \ga \ln \psi + f(t) \right\} \psi   
\end{eqnarray}
and defining the diffusion coefficient after Nelson \cite{Ne-PR-1966} as 
\begin{eqnarray} \label{eq: D}
	\mathcal{D} &=&  \frac{\hb}{2m}   ,
\end{eqnarray}
Eq. (\ref{eq: eq1}) leads to a generalized nonlinear Schr\"odinger equation 
\begin{eqnarray} \label{eq: gen_SCH1}
	i \hb \frac{\pa}{\pa t} \psi &=& \left\{
	-\frac{\hb^2}{2m}  \frac{\pa^2}{\pa x^2}  + V + V_r - i \ga^{}_R \hb ( \ln\psi - \la \ln\psi \ra )+ \hb \ga^{}_I ( \ln\psi - \la \ln\psi \ra ) \right\} \psi \nonumber \\
\end{eqnarray}
where $\ga^{}_R$ and $ \ga^{}_I $ are the real and imaginary parts of the complex friction coefficient $\ga$, 
playing the role of {\it classical} and 
{\it quantum} friction coefficients, respectively. The imaginary part has been related to an effective temperature $T_e$ 
(positive or negative) through $2 k_B T_e / \hbar$ where $k_B$ is the Boltzmann constant \cite{Ch-EPJP-2017}. 
The term accompanying $\gamma_I$ in Eq. (\ref{eq: gen_SCH1}) can then be considered as a statistical potential  leading  under certain 
conditions  to the Gross-Pitaevskii equation \cite{Ch-EPJP-2017}. This is a reminiscent of 
the symmetry of the wave function describing an  ensemble of particles (bosons and fermions) as the origin of what 
is called statistical potential \cite{Pathria}, which can be attractive or repulsive. Finally, the integration constant $f(t)$ is set in such a 
way that the expectation value of the friction term proportional to $\ga^{}_R$ is zero.

From the polar form of the wave function Eq. (\ref{Psi}),
%
%
Eq. (\ref{eq: gen_SCH1}) can be split again into real and imaginary parts according to \cite{Mousavi-Salva-2019-1}
\begin{align}~ 
	-\frac{ \pa S }{ \pa t } = \frac{1}{2m} \left(\frac{\pa S}{\pa x} \right)^2 + V + V_r + Q + \ga^{}_R (S - \la S \ra) + \frac{\hb \ga^{}_I}{2} ( \ln \rho - \la \ln \rho \ra ) \label{eq: QHJ1}
	\\
	\frac{ \pa \rho }{ \pa t } + \frac{\pa }{\pa x} \left( \rho \frac{1}{m} \frac{\pa S}{\pa x} \right) = - \ga^{}_R \rho ( \ln \rho - \la \ln \rho \ra ) + \frac{2 \ga^{}_I }{\hb} \rho ( S - \la S \ra )
	\label{eq: con1}
\end{align}
where 
\begin{eqnarray}
\rho(x, t) &=& \lvert \psi(x, t) \rvert^2  ,  \\
Q(x, t) & = & - \frac{\hb^2}{2m} \frac{1}{\sqrt{\rho}} \frac{\pa^2 \sqrt{\rho} }{\pa x^2}   .
\end{eqnarray}
Eqs. (\ref{eq: QHJ1}) and (\ref{eq: con1}) can now be seen as the generalized Hamilton-Jacobi and continuity equations, respectively.
This  continuity equation has now two source/sink terms and therefore Eq. (\ref{eq: gen_SCH1}) violates the local conservation 
of probability density. The integration over the whole space of Eq. (\ref{eq: con1}) provides however the correct global conservation 
of the normalization. The partial derivative  with respect to the space coordinate  of Eq. (\ref{eq: QHJ1}) yields 
\begin{eqnarray} \label{eq: nabla_QHJ1}
	\left( \frac{ \pa }{ \pa t } + \frac{1}{m} \frac{\pa S}{\pa x} \frac{\pa }{\pa x} \right) \frac{\pa S}{\pa x}
	&=& - \frac{\pa }{\pa x} \left( V + V_r + Q  \right)  - \ga^{}_R \frac{\pa S}{\pa x} - \frac{\hb \ga^{}_I}{2 \rho} \frac{\pa \rho}{\pa x} .
\end{eqnarray}

At this point it is of interest to discuss Ehrenfest's theorem for complex friction.
As is well known, Ehrenfest's relations in three dimensions are given by 
\begin{align}~ 
	\frac{d}{dt} \la \hat{ \mathbf{r} } \ra = \frac{ \la \hat{ \mathbf{p} } \ra }{m}   , \label{eq: Eh1} 
	\\
	\frac{d}{dt} \la \hat{ \mathbf{p} } \ra = - \la \nabla V \ra - \ga^{}_R \la \hat{ \mathbf{p} } \ra   . \label{eq: Eh2}
\end{align}
%
For interaction potentials of at most second order in space coordinates where $ \la \nabla V(\mathbf{r}) \ra = \nabla V(\la \mathbf{r} \ra) $, 
after Eq. (\ref{eq: Eh1}), Eq. (\ref{eq: Eh2}) can be written as 
\begin{eqnarray}\label{eq: Eh_theor}
	m \frac{d^2}{dt^2} \la \hat{ \mathbf{r} } \ra &=& - \nabla V(\la \mathbf{r} \ra) - \ga^{}_R \la \hat{ \mathbf{p} } \ra   ,
\end{eqnarray} 
which is the classical equation of motion for the expectation value of the position operator. This equation is known as Ehrenfest's theorem
\cite{BaYaZi-PRA-1994}. One can see that the standard continuity equation (the continuity equation without source/sink terms), which preserves 
local conservation of normalization, is a {\it sufficient} condition for fulfilling Eq. (\ref{eq: Eh1}).

However, the generalized nonlinear Schr\"{o}dinger equation (\ref{eq: gen_SCH1}) violates both Ehrenfest relations.
The time derivative of the expectation value of the position operator is given by
\begin{eqnarray} \label{eq: d<r>_dt}
	\frac{d}{dt} \la \hat{ \mathbf{r} } \ra &=& \int d^3x  \frac{\pa \rho}{\pa t} \mathbf{r}
	= \frac{ \la \hat{ \mathbf{p} } \ra }{m} - \ga^{}_R \big \la \hat{ \mathbf{r} } ( \ln \rho - \la \ln \rho \ra )  \big \ra
	+ \frac{2\ga^{}_I}{\hb} \big \la \hat{ \mathbf{r} } ( S - \la S \ra )  \big \ra  ,
\end{eqnarray}
where  integration by parts and  the relation $ \la \hat{ \mathbf{p} } \ra = \la \nabla S \ra $ have been applied. 
Eq. (\ref{eq: d<r>_dt}) explicitly shows violation of Eq. (\ref{eq: Eh1}). 
%
%
Analogously, the same happens with Eq. (\ref{eq: Eh2}). For this goal, we multiply both sides of Eq. (\ref{eq: nabla_QHJ1}) 
by $\rho$. Then, by taking into account the continuity equation (\ref{eq: con1}), one has that
\begin{equation} \label{eq: aux2}
	\begin{aligned}
		& \frac{\pa ( \rho \nabla S )}{\pa t} + \sum_i \mathbf{e}_i \nabla \cdot \left( \rho \frac{\nabla_i S}{m} \nabla S \right) + \nabla S \left( \ga^{}_R \rho( \ln\rho - \la \ln\rho \ra) - \frac{2\ga_I}{\hb} \rho ( S - \la S \ra )   \right) 
		\\ 
		& = - \rho \nabla(V + Q) - \ga^{}_R \rho \nabla S - \frac{\hb \ga^{}_I}{2} \nabla \rho   ,
	\end{aligned}
\end{equation}
where $ \mathbf{e}_i $ denotes the unit vector along the $x_i$ direction .
Now by integrating both sides of this equation over the whole space and given that $ \la \nabla Q \ra = 0 $, one has that
\begin{eqnarray} \label{eq: ehren}
	\frac{d}{d t} \la \nabla S \ra &=& - \la \nabla V \ra 
	- \ga^{}_R \la \nabla S \ra
	- \ga^{}_R \bigg \la \nabla S ( \ln\rho - \la \ln\rho \ra )  \bigg \ra
	+ \frac{2\ga^{}_I}{\hb} \bigg \la  \nabla S ( S - \la S \ra )  \bigg \ra . \nonumber \\
\end{eqnarray}
%
For solutions where the phase of the wave function is linear in space, $\nabla S$ is only a function of time. Thus, the last two terms of 
right hand side of Eq. (\ref{eq: ehren}) becomes zero. Only in this case, this Ehrenfest relation is fulfilled.

\subsection{The Nassar-Miret-Art\'es equation}\label{subsec2-4}

An even more general nonlinear Schr\"odinger equation was proposed by Nassar and Miret-Art\'es (NMA) \cite{NaMi-PRL-2013}
to describe a stochastic dynamics together with continuous measurements
\begin{eqnarray}  \label{eq: NM}
	i \hb \frac{\pa}{\pa t}\psi &=& \left[ -\frac{\hb^2}{2m} \nabla^2
	+ V + V_r +  i \hb \, ( W_{\text{c}} + W_{\text{f}} )
	\right] \psi(x, t)   , \nonumber \\
\end{eqnarray}
where
\begin{eqnarray}  
	W_{\text{c}}(x, t) &=& -\kappa ( \ln \vert \psi \vert^2 - \langle \ln \vert \psi \vert^2  \rangle )   , \label{eq: pot_Wc} \\
	W_{\text{f}}(x, t) &=& -\frac{\ga}{2} \left(  \ln \frac{\psi}{\psi^*} - \left\langle  \ln \frac{\psi}{\psi^*} \right\rangle \right) , \label{eq: pot_Wf}
\end{eqnarray}
with real parameters $\ga$ and $\kappa$. Here, $ \kappa $ plays the role of the resolution of the continuous measurement apparatus.  
If the polar form of the wave function Eq. (\ref{Psi}) 
is again substituted into Eq. (\ref{eq: NM}), the resulting equations for the real and imaginary parts are expressed now as
\begin{align}~ 
	-\frac{\pa S}{\pa t} = \frac{1}{2m} \bigg(\frac{\partial S}{\partial x}\bigg)^2 + V + V_r(x,t) + Q + \ga ( S - \langle S \rangle ) , \label{eq: HJ_NM} 
	\\
	\frac{\pa \rho }{\pa t} + \frac{\partial }{\partial x} \left( \rho \frac{1}{m}\frac{\partial S}{\partial x} \right) = - 2 \kappa \rho ( \ln \rho - \langle \ln \rho \rangle )  . \label{eq: con_NM}
\end{align}
Comparison of Eq. (\ref{eq: QHJ1}) with Eq. (\ref{eq: HJ_NM}) and Eq. (\ref{eq: con1}) with Eq. (\ref{eq: con_NM}) shows that when 
$\ga^{}_I = 0 $ and $ \ga^{}_R = 2 \kappa $, the generalized nonlinear  Schr\"{o}dinger equation (\ref{eq: gen_SCH1}) is equivalent to the NMA
equation (\ref{eq: NM}) for the special value $ \kappa = \ga / 2 $. 
%

%
%

In the previous section, we have showed that Eq. (\ref{eq: gen_SCH1}) violates local conservation of the 
probability density function. 
In this context, another generalized differential equation has been proposed \cite{Ch-EPJP-2017}  
\begin{eqnarray} \label{eq: gen_SCH2}
	i \hb \frac{\pa \psi}{\pa t} &=& \bigg\{
	-\frac{\hb^2}{2m} \frac{\partial^2}{\partial x^2} + V + V_r + \hb \ga^{}_I \ln(\vert \psi \vert) \nonumber \\
	&&+ \frac{\hb}{2i} \ga^{}_R \left[ \ln \left( \frac{\psi}{\psi^*} \right)  - \left \la \ln \left( \frac{\psi}{\psi^*} \right)  \right \ra  \right] \bigg\} \psi     . 
\end{eqnarray}
As mentioned above, Chavanis \cite{Ch-EPJP-2017} introduced  $\ga^{}_I = 2 k_B T_e /\hbar$ 
leading to a form of fluctuation-dissipation theorem between real and imaginary 
parts of the friction coefficient, $T_e$ being an effective temperature.
If $ \ga^{}_I=0 $, Eq. (\ref{eq: gen_SCH2}) reduces to the Kostin or SL equation (\ref{eq: Sch_Lan}). For an imaginary friction coefficient, it 
reduces to BBM equation \cite{BBMy-AP-1976} which possesses soliton-like solutions of Gaussian shape. 
Eq. (\ref{eq: gen_SCH2}) can then also be proposed to describe continuous measurements in dissipative media as an alternative to the 
NMA equation.

Once more from the polar form of the wave function (\ref{Psi}), Eq. (\ref{eq: gen_SCH2}) leads to
\begin{equation} 
	-\frac{\pa S }{ \pa t } = \frac{1}{2m} \bigg(\frac{\partial S}{\partial x}\bigg)^2  + V + V_r + \ga^{}_R (S - \la S \ra) - \frac{\hbar^2}{2m} \frac{1}{ \vert \psi \vert  }  
	  \frac{\partial^2 \vert \psi \vert}{\partial x^2}   + \hb \ga^{}_I \ln \vert \psi \vert  	\label{eq: QHJ2}
\end{equation} 
and
\begin{equation} 
\frac{ \pa \rho }{ \pa t } +  \frac{\partial }{\partial x}  \left( \rho \frac{1}{m}  \frac{\partial S}{\partial x} \right) =	0     .
	\label{eq: con2}
\end{equation} 
These equations are the modified Hamilton-Jacobi and continuity equations, respectively. The space derivative of Eq. (\ref{eq: QHJ2}) is
\begin{equation} \label{eq: der_QHJ2}
	\left( \frac{ \pa }{ \pa t } + \frac{1}{m}  \frac{\partial S}{\partial x}  \frac{\partial }{\partial x} \right)  \frac{\partial S}{\partial x}
	= -  \frac{\partial }{\partial x} \bigg( V + V_r - \frac{\hb^2}{2m} \frac{1}{ \lvert \psi \rvert }
	\frac{\partial^2 \vert \psi \vert}{\partial x^2}  + \hb \ga^{}_I \ln \vert \psi \vert \bigg)  - \ga^{}_R  \frac{\partial S}{\partial x} 
\end{equation}
and comparison with the classical equation of motion suggests us to define a (Bohmian) velocity field as \cite{Holland-book-1993}  
\begin{eqnarray} \label{eq: BMvel2}
v(x, t) &=& \frac{1}{m}  \frac{\partial S}{\partial x}  ,
\end{eqnarray}
from which Eq. (\ref{eq: der_QHJ2}) can be recast
\begin{eqnarray} \label{eq: motion2}
	\frac{ d v}{ d t } &=& - \frac{1}{m} \frac{\partial (V + Q)}{\partial x} - \ga^{}_R v + \frac{1}{m} F_r(t) - \frac{\hb \ga^{}_I}{2m} \frac{\partial }{\partial x}  \ln \rho   .
\end{eqnarray}
Thus,  the new term $ \frac{\hb \ga^{}_I}{2} \ln \rho $ is actually an 
additional contribution to  $Q$. Due to the fact that Bohmian particles are distributed according to the Born rule, 
Bohmian average values of physical quantities are just quantum expectation values.
Let us see an illustration. The energy of a Bohmian particle is given by
\begin{eqnarray} \label{eq: En2}
E(x, t) &=& \frac{1}{2m} \bigg(\frac{\partial S}{\partial x}\bigg)^2 + V + V_r + Q + \frac{\hb \ga^{}_I}{2} \ln \rho = -\frac{ \pa S }{ \pa t } - \ga^{}_R (S - \la S \ra)  ,
\end{eqnarray}
where Eq. (\ref{eq: QHJ2}) has been used in the second part of the equation. Thus, its average is expressed as
\begin{eqnarray} \label{eq: expEn2}
	\la E \ra &=& \int d x ~ \vert \psi(x, t) \vert^2 E(x, t) \nonumber \\
	&=& \frac{1}{2m} \int d x ~ \bigg(\vert \psi \vert^2 \bigg( \frac{\partial S}{\partial x}\bigg)^2 + \hb^2  \big\vert \frac{\partial  \psi }{\partial x} \big\vert^2 \bigg)+  \int dx ~ \vert \psi \vert^2 (V + V_r + \hb \ga^{}_I \ln \vert \psi \vert) \nonumber  \\
	&=& \frac{\la \hat{p^2 \ra }}{2m} + \la V \ra + \la V_r \ra + \frac{\hb \ga^{}_I}{2} \la \ln \rho \ra    .
\end{eqnarray}
The energy expectation value is then calculated by \cite{Ko-JCP-1972} 
\begin{eqnarray} \label{eq: expEn2}
	\la E \ra &=& \int dx ~ \psi^* i\hb \frac{\pa}{\pa t} \psi
\end{eqnarray}
and the same result is reached. Now, if we recall that for any arbitrary function $A(x, t)$ 
\begin{eqnarray} \label{eq: property}
	\frac{d}{dt} \la A \ra &=& \int dx \frac{\pa}{\pa t} (A \vert \psi \vert^2) =  \left \la \frac{dA}{dt} \right\ra  ,
\end{eqnarray}
where Eq. (\ref{eq: con2}) and an integration by parts have been used 
%
then, from Eq. (\ref{eq: motion2}) and Eqs. (\ref{eq: En2}) and 
(\ref{eq: property}); and the fact that $ \la \pa Q / \pa t  \ra = 0$ and $\la \pa \ln \rho / \pa t  \ra =0$, the time variation of the energy is finally given by  \cite{Ko-JPA-2007}
\begin{eqnarray} \label{eq: en_rate}
	\frac{d}{dt} \la E \ra &=& - \ga^{}_R \frac{1}{m} \left \la \bigg(\frac{\partial S}{\partial x}\bigg)^2 \right\ra + \left \la \frac{\pa V}{\pa t} \right\ra + \left \la x \frac{d F_r(t)}{dt} \right\ra . 
\end{eqnarray}

\section{Dissipative and Stochastic Scaled trajectories}\label{sec3}

\subsection{The Caldirola-Kanai equation}\label{subsec3-1}

Following Rosen \cite{Rozen}, by subtracting the non-linear term $ e^{-2\gamma t} Q $ to the 
classical potential term $ V $ in the CK equation (\ref{eq: Sch_viscid}), 
the classical wave equation in the CK framework is written as
\begin{eqnarray}
	i \hbar \frac{\partial}{\partial t}\psi_{\text{cl}}(x, t) &=&  \bigg[ -\frac{\hbar^2}{2m} e^{-\gamma t}
	\frac{\partial^2}{\partial x^2} + V(x) e^{\gamma t}  \nonumber \\
	&&+ \frac{\hbar^2}{2m} \frac{e^{-\gamma t}}{\vert \psi_{\text{cl}}(x, t) \vert} \frac{\partial^2 \vert \psi_{\text{cl}}(x, t) \vert}{ \partial x^2}   
	\bigg] \psi_{\text{cl}}(x, t) .  \label{eq: class_wave equation}
\end{eqnarray}
%
%
This  classical nonlinear differential equation (\ref{eq: class_wave equation}) can be seen as the limiting case 
of the linear equation (\ref{eq: Sch_viscid}). The last term in (\ref{eq: class_wave equation}) which is 
proportional to the quantum potential is responsible for the nonlinearity and important to recover the classical Hamilton-Jacobi equation.
The wave-particle duality is thus present in this classical wave differential equation but the wave function does not play the role 
of Bohmian mechanics, the state of a classical system is still determined by its position and momentum.
%
After Holland \cite{Holland-book-1993}, one {\it continuously} passes from one extreme regime to the other by varying 
the effectiveness of the quantum potential. A similar approach is also used by Allori et al. 	\cite{Allori-2002}.

An alternative way to establish a continuous quantum-classical transition is through what is known the quantum-classical transition 
wave differential equation according to  \cite{RiSchMaVaBa-PRA-2014}
\begin{eqnarray}
	i \hbar \frac{\partial}{\partial t}\psi_{\epsilon}(x, t) &=& 
	\bigg[ -\frac{\hbar^2}{2m} e^{-\gamma t}
	\frac{\partial^2}{\partial x^2} + V(x) e^{\gamma t} \nonumber \\
	 && +(1-\epsilon)  \frac{\hbar^2}{2m} \frac{e^{-\gamma t}}{\vert \psi_{\epsilon}(x, t) \vert} \frac{\partial^2 \vert \psi_{\epsilon}(x, t) \vert}{ \partial x^2}   
	\bigg] \psi_{\epsilon}(x, t) ,  \label{eq: ck_quan_class transition}
\end{eqnarray}
where a degree of quantumness ruled by the transition parameter $\epsilon$ (with $ 0 \leq \epsilon \leq 1 $) has 
been included. This transition differential equation provides a {\it continuous} description of the dissipative process of a physical
system from going to one extreme regime to the other and in-between. Furthermore, for  $\epsilon=0$, 
this transition equation reduces to the classical wave equation (\ref{eq: class_wave equation}) 
while, for $\epsilon=1$, it reduces to the Schr\"{o}dinger equation (\ref{eq: Sch_viscid}) in the
CK framework.

Following the same procedure, by substituting the polar form of the wave function written now as
\begin{equation}\label{Psie}
\psi_{\epsilon} (x,t)=R_{\epsilon}(x, t) e^{i S_{\epsilon}(x, t)/ \hbar }
\end{equation}
into Eq. (\ref{eq: ck_quan_class transition}) and after some straightforward manipulations, 
the following equations are reached 
\begin{eqnarray}
	- \frac{\partial  S_{\ep}}{\partial t} \ti{\psi} &=& \frac{1}{2m} e^{-\gamma t} \left( \frac{\partial  S_{\ep}}{\partial x} \right)^2  \ti{\psi} + V(x) e^{\gamma t} \ti{\psi} -  \frac{ \ti{\hbar}^2}{2m} e^{-\gamma t} \frac{1}{R_{\ep}}
	\frac{\partial^2  R_{\ep}}{\partial x^2} \ti{\psi} , \nonumber \\ \label{eq: tran_real_part2}
	\\
	i \ti{\hbar} \frac{\partial  R_{\ep}}{\partial t} e^{iS_{\ep}/\ti{\hbar}} &=&  -\frac{\ti{\hbar}^2}{2m } e^{-\gamma t}  \left[ \frac{2i}{\ti{\hbar}}
	\frac{\partial  R_{\ep}}{\partial x} \frac{\partial  S_{\ep}}{\partial x} e^{ iS_{\ep}/\ti{\hbar} } + \frac{i}{\ti{\hbar}} \frac{\partial^2  S_{\ep}}{\partial x^2} \ti{\psi} \right] , \label{eq: tran_imag_part2}
\end{eqnarray}
where the scaled Planck constant as well as the scaled wave function in polar form have been defined as
\begin{eqnarray} \label{eq: scaled Planck}
	\ti{\hbar} &=& \hbar ~ \sqrt{\epsilon} ,
\end{eqnarray}
and
\begin{eqnarray} \label{eq: scaled_wf_polar}
	\ti{\psi}(x, t) &=& R_{\ep}(x, t) e^{i S_{\ep}(x, t)/ \ti{\hbar} },
\end{eqnarray}
respectively.
Now, by adding Eq. (\ref{eq: tran_real_part2}) and Eq. (\ref{eq: tran_imag_part2}),  the scaled linear 
Schr\"{o}dinger equation is obtained
\begin{eqnarray} \label{eq: Scaled CK}
	i \tilde{\hbar} \frac{\partial}{\partial t}\ti{\psi}(x, t) &=& \left[ -\frac{\tilde{\hbar}^2}{2m} e^{-\gamma t}
	\frac{\partial^2}{\partial x^2} + V(x) e^{\gamma t}
	\right] \ti{\psi}(x, t) .
\end{eqnarray}
In other words, the nonlinear transition differential equation (\ref{eq: ck_quan_class transition}) is equivalent to the 
scaled linear Schr\"{o}dinger equation (\ref{eq: Scaled CK}) and similar to 
Eq. (\ref{eq: Sch_viscid}). 
The transition wave function is expressed as
\begin{eqnarray} \label{eq: scaled_transtion relation}
	\ti{\psi}(x, t) &=& \psi_{\ep}(x, t) \exp \left[ \frac{i}{\hbar} \left( \frac{1}{\sqrt{\ep}} - 1  \right) S_{\ep}(x, t) \right] .
\end{eqnarray}
Dissipative scaled trajectories are issued from Eq. (\ref{eq: Scaled CK}).

\subsection{The Schr\"odinger-Langevin equation}\label{subsec3-2}

Following the same procedure as before \cite{RiSchMaVaBa-PRA-2014},  
Eq. (\ref{eq: Sch_Lan}) can be rewritten as a quantum-classical transition wave differential equation  as follows 
\begin{eqnarray} 
	i \hbar \frac{\partial}{\partial t}\psi_{\ep}(x, t) &=& \bigg[ -\frac{\hbar^2}{2m}
	\frac{\partial^2}{\partial x^2} + V(x, t) + V_r(x, t) 
	+ \frac{\gamma \hbar}{2 i} \left( \ln \frac{\psi_{\ep}}{\psi_{\ep}^*} - \left \langle \ln \frac{\psi_{\ep}}{\psi_{\ep}^*} \right \rangle \right) 
	\nonumber \\
	& &~~+ (1-\ep) \frac{\hbar^2}{2m} 
	\frac{1}{\vert \psi_{\epsilon}(x, t) \vert} 
	\frac{\partial^2 \vert \psi_{\epsilon}(x, t) \vert}{\partial x^2 } 
	\bigg ] \psi_{\ep}(x, t) ,  \label{eq: quan_class transition}
\end{eqnarray}
where the degree of quantumness is given again by the transition parameter. The wave function 
$\psi_{\ep}(x, t)$ is thus affected by this parameter ruling each open dynamical regime.
By substituting now the polar form of this wave function, given by Eq. (\ref{Psie}), into Eq. (\ref{eq: quan_class transition}), 
the following coupled equations for the amplitude $R_{\ep}(x,t)$ and phase $S_{\ep}(x, t)$  are reached
\begin{eqnarray}
	\frac{\partial  R_{\ep}}{\partial t} &=&  -\frac{1}{2m } \left( 2 \frac{\partial  R_{\ep}}{\partial x} \frac{\partial  S_{\ep}}{\partial x} + R_{\ep} \frac{\partial^2  S_{\ep}}{\partial x^2} \right) , \label{eq: tran_imag_part}
	\\
	- \frac{\partial  S_{\ep}}{\partial t} R_{\ep} &=& - \frac{\hbar^2}{2m} \left[  \ep \frac{\partial^2  R_{\ep}}{\partial x^2}
	- \frac{1}{\hbar^2} R_{\ep} \left( \frac{\partial  S_{\ep}}{\partial x} \right)^2
	\right ] + \bigg[ V + V_r + \gamma ( S_{\ep} - \langle S_{\ep} \rangle  )  \bigg]  R_{\ep} , \nonumber \\
	\label{eq: tran_real_part}
\end{eqnarray}
with
\begin{eqnarray} \label{eq: tran_wave_phase}
	S_{\ep} &=& \frac{\hbar}{2i} \ln \frac{\psi_{\ep}}{\psi_{\ep}^*} .
\end{eqnarray}
Furthermore, by using again the  scaled Plank's constant and scaled wave function and substituting  into Eqs. (\ref{eq: tran_imag_part}) 
and  (\ref{eq: tran_real_part})  the scaled nonlinear SL equation for a stochastic dynamics is expressed as 
\begin{eqnarray} \label{eq: Scaled Sch-Lan-1}
	i \ti{\hbar} \frac{\partial}{\partial t}\ti{\psi}(x, t) &=& \left[ -\frac{\ti{\hbar}^2}{2m} \frac{\partial^2}{\partial x^2}
	+ V + V_r + \frac{\gamma \ti{\hbar}}{2 i} \left( \ln \frac{\ti{\psi}}{\ti{\psi}^*} - \left \langle \ln \frac{\ti{\psi}}{\ti{\psi}^*} \right \rangle \right) 
	\right] \ti{\psi}(x, t)   . \nonumber \\
\end{eqnarray}
Stochastic scaled trajectories are issued from this nonlinear equation.
When the random potential $V_r(x,t)$ is neglected, the dissipative system is described by 
\begin{eqnarray} \label{eq: Scaled Sch-Lan}
	i \ti{\hbar} \frac{\partial}{\partial t}\ti{\psi}(x, t) &=& \left[ -\frac{\ti{\hbar}^2}{2m} \frac{\partial^2}{\partial x^2}
	+ V +  \frac{\gamma \ti{\hbar}}{2 i} \left( \ln \frac{\ti{\psi}}{\ti{\psi}^*} - \left \langle \ln \frac{\ti{\psi}}{\ti{\psi}^*} \right \rangle \right) 
	\right] \ti{\psi}(x, t)   ,
\end{eqnarray}
being again a nonlinear logarithmic equation. As mentioned above, the resulting equation is an alternative description of the dissipative 
dynamics compared to the CK view. In any case, Eq. (\ref{eq: quan_class transition}) is equivalent to Eq.
(\ref{eq: Scaled Sch-Lan-1}) (or  Eq. (\ref{eq: Scaled Sch-Lan}) only for the dissipative case).  
Moreover, the corresponding wave functions and phases are related by Eq. (\ref{eq: scaled_transtion relation}).
%
%

The same procedure could be again followed for the next two nonlinear differential equations above mentioned but we are not going 
to show them here.
Thus, the decoherence process resulting from both the open quantum dynamics  and scaled Planck's
constant is carried out in a gradual way within each dynamical regime (it is worth mentioning that the environment is also seen as acting like 
a continuous measuring apparatus).

\section{Propagation of Gaussian wave packets. The generalized Penney equation} \label{sec4}

Let us consider a time-dependent Gaussian ansatz for the probability density \cite{NaMi-book-2017}
\begin{eqnarray} \label{eq: rho_ansatz}
	\rho(x, t) &=& \frac{1}{\sqrt{2\pi}\si(t)} \exp \left[ -\frac{(x-q(t))^2}{2 \si^2(t)} \right] ,
\end{eqnarray}
where $q(t) = \int dx ~ x \, \rho(x, t) $ is the expectation value of the position operator.  Thus, the center of the corresponding Gaussian 
wave packet follows in principle the classical trajectory $q(t)$ ( see however Eq. (78)) and its time dependent width is $\si(t)$. 
For a dissipative motion where $V_r =0$,
if we first divide Eq. (\ref{eq: con1}) by $\rho$ and then derive the resulting equation with respect to $x$, we have that
\begin{eqnarray} \label{eq: aux1}
	\frac{\pa^3 S}{\pa x^3}&+& \frac{1}{\rho} \frac{\pa \rho}{\pa x} \frac{\pa^2 S}{\pa x^2} + 
	\bigg[ - \frac{2m\ga_I}{\hb} - \frac{1}{\rho^2} \left(\frac{\pa \rho}{\pa x} \right)^2 + \frac{1}{\rho} \frac{\pa^2 \rho}{\pa x^2} \bigg] \frac{\pa S}{\pa x} + \nonumber \\
	&+& \frac{m}{\rho} \bigg( \ga^{}_R \frac{\pa \rho}{\pa x} - \frac{1}{\rho} \frac{\pa \rho}{\pa x} \frac{\pa \rho}{\pa t} + \frac{\pa^2 \rho}{\pa x \pa t}
	\bigg) =0    .
\end{eqnarray}
If one assumes the ansatz
\begin{eqnarray} \label{eq: pa_xS ansatz}
	\frac{\pa S}{\pa x} &=& a(t) ( x - q(t) ) + b(t) ,
\end{eqnarray}
and introduces Eqs. (\ref{eq: rho_ansatz}) and (\ref{eq: pa_xS ansatz}) into Eq. (\ref{eq: aux1}), the corresponding time dependent 
coefficients $a(t)$ and $b(t)$ are given by
\begin{eqnarray} 
	a(t) &=& m \frac{ \dot{\si}(t) - \frac{\ga^{}_R}{2} \si(t) }{\si(t) \left( 1 + \frac{m \ga^{}_I}{\hb}\si(t)^2 \right) }  , \label{eq: at} \\
	b(t) &=& m \frac{ \dot{q}(t)}{ 1 + \frac{2 m \ga^{}_I}{\hb}\si(t)^2 }  , \label{eq: bt}
\end{eqnarray}
where the linear independence of different powers of $(x - q(t))$ has been considered. Thus, one has now that
\begin{eqnarray} \label{eq: S}
	S(x, t) &=& \frac{m}{2} \frac{ \dot{\si}(t) - \frac{\ga^{}_R}{2} \si(t) }{\si(t) \left( 1 + \frac{m \ga^{}_I}{\hb}\si(t)^2 \right) } ( x - q(t) )^2 + m \frac{ \dot{q}(t)}{ 1 + \frac{2 m \ga^{}_I}{\hb}\si(t)^2 } ( x - q(t) ) + g(t) \nonumber \\
\end{eqnarray}
where $g(t)$ is the constant of integration which can be determined by Eq. (\ref{eq: QHJ1}).
The explicit expressions for $q(t)$ and $\si(t)$ are obtained 
by introducing Eq. (\ref{eq: S}) into Eq. (\ref{eq: nabla_QHJ1}), 
expanding the interaction potential $V(x)$ around $q(t)$ up to second order, and solving  the following differential equation
for the width
\begin{equation} \label{eq: sigma(t)}
	\begin{aligned}
		& \left( 1 + \frac{m \ga^{}_I \si(t)^2}{\hb} \right) \ddot{\si}(t)
		- \frac{3 m \ga^{}_I \si(t)}{\hb} \dot{\si}(t)^2
		+ \frac{2 m \ga^{}_I \ga^{}_R \si(t)^2}{\hb}\dot{\si}(t)
		+ \frac{m V_2 \ga_I^2}{\hb^2}\si(t)^5  \\
		& + \frac{\ga^{}_I}{2 \hb} \left( 4 V_2 - m \ga_I^2 - m \ga_R^2 \right) \si(t)^3
		+ \left(  \frac{V_2 }{m} - \frac{\ga_R^2 + 5 \ga_I^2}{4} \right) \si(t)
		- \frac{\ga^{}_I \hb}{m \si(t)} - \frac{\hb^2}{4 m^2 \si(t)^3} = 0  ,
	\end{aligned}
\end{equation}
and for the center of the wave packet
\begin{equation} \label{eq: q(t)}
	\begin{aligned}
		& \left[ 1 + \ga^{}_I \left( \frac{3 m \si(t)^2}{\hb} + \frac{2 m^2 \ga^{}_I \si(t)^4}{\hb^2} \right) \right] \ddot{q}(t) \\
		& + \left[ \ga^{}_R + 
		\ga^{}_I \left( \frac{4 m  \ga^{}_R \si(t)^2}{\hb} + \frac{4 m^2 \ga^{}_I \ga^{}_R \si(t)^4}{\hb^2} 
		- \frac{6 m \si(t) \dot{\si}(t)}{\hb} - \frac{8 m^2 \ga^{}_I \si(t)^3 \dot{\si}(t)}{\hb^2} 
		\right)
		\right] \dot{q}(t) \\
		& = - \frac{V_1}{m} \left[ 1 + \ga^{}_I \left(\frac{5 m \si(t)^2}{\hb} + \frac{8 m^2 \ga^{}_I \si(t)^4}{\hb^2} 
		+ \frac{4 m^3 \ga_I^2 \si(t)^6}{\hb^3}
		\right)\right]   ,
	\end{aligned}
\end{equation}
where
\begin{eqnarray} 
	V_1 &=& \frac{\pa V}{\pa x}  \bigg\vert_{x=q(t)} ~ \mbox{and} ~~~~
	V_2 = \frac{\pa^2 V}{\pa x^2} \bigg\vert_{x=q(t)}   .
\end{eqnarray}
Again, the linear independency of different powers of $(x - q(t))$ has been used.
This procedure is exact for potentials of at most second order in space coordinates and one should expect that the center 
of the wave packet follows a classical trajectory. Eq. (\ref{eq: q(t)}) explicitly shows however that, in general,  $q(t)$ does not follow  
a classical trajectory since the width is involved in its equation of motion. Thus, for complex friction coefficients, the dressing 
scheme mentioned above is no longer valid. 
For the special case of a  real friction coefficient, $q(t)$ do follow the classical equation of motion. 
Furthermore, the differential equation for the width is the well-known dissipative Pinney equation \cite{NaMi-book-2017}.  In fact, 
Eq. (\ref{eq: sigma(t)}) can be seen as a generalized dissipative Pinney equation.  Dissipative Bohmian trajectories are issued 
from these two time differential equations.

For stochastic Bohmian trajectories, $V_r \neq 0$, one has to solve Eqs. (\ref{eq: con2}) and (\ref{eq: motion2}) together with
the time-dependent Gaussian ansatz (\ref{eq: rho_ansatz}) for the probability density.
This Gaussian ansatz satisfies Eq. (\ref{eq: con2}) provided that the velocity field and trajectories are given by
\begin{eqnarray} \label{eq: BM_vel_Gauss}
	v(x, t) &=& \frac{{\dot{\si}(t)}}{\si(t)} (x-q(t)) + \dot{q}(t) 
\end{eqnarray}
and 
\begin{eqnarray} \label{eq: BM-trajs} 
	x(x^{(0)}, t) &=& q(t) + (x^{(0)} - q(0)) \frac{\si(t)}{\si(0)} ,
\end{eqnarray}
where $x^{(0)}$ and $q(0)$ are the initial conditions for $x(x^{(0)}, t)$ and $q(t)$, respectively.
As mentioned above, Eqs.  (\ref{eq: BM_vel_Gauss}) and (\ref{eq: BM-trajs}) display the standard dressing scheme for the 
velocity and position \cite{NaMi-book-2017}.

By introducing again the ansatz (\ref{eq: rho_ansatz}) and Eq. (\ref{eq: BM_vel_Gauss}) into the equation of motion (\ref{eq: motion2}), one obtains
\begin{align}~
	\ddot{q}(t) + \ga^{}_R \dot{q}(t) + \frac{1}{m} \frac{\pa V}{\pa x}  \bigg\vert_{x=q(t)} = F_r(t) , \label{eq: xbar}  \\
	\ddot{\si}(t) + \ga^{}_R \dot{\si}(t) - \frac{\hb^2}{4 m^2 \si(t)^3} - \frac{\hb \ga^{}_I }{2m}  \frac{1}{\si(t)} + \si(t) \frac{1}{m} \frac{\pa^2 V}{\pa x^2} \bigg\vert_{x=q(t)} = 0  \label{eq: delta} 
\end{align}
where the interaction potential around the classical path $q(t)$ up to second order has been expanded. 
As can be seen, $q(t)$ is governed by the classical Langevin equation of motion and $\sigma(t)$ by the generalized Pinney equation; 
the new term with $\gamma_I$ provides the extension of the standard dissipative Pinney equation \cite{NaMi-book-2017,Mousavi-Salva-2019-1}.
The equation is also known as Ermakov equation and appears, for example, in the process of cooling down atoms in a harmonic 
trap \cite{Mu-PRL}. It should be mentioned that the differential
equations fulfilled by $q(t)$ and $\sigma (t)$ are not coupled each other; the width is not influenced by the random force. 
Chavanis \cite{Ch-EPJP-2017} pointed out that in cosmology,  the so-called Hubble parameter which can be defined as the ratio 
$\dot \sigma / \sigma$, the corresponding width also follows Eq. (\ref{eq: delta}). 
For time-independent quadratic potentials, there is a soliton-like solution for Eq. (\ref{eq: delta}) with $ \dot{\si} = 0 $,
\begin{eqnarray} \label{eq: soliton}
	\ga^{}_I &=& - \frac{\hb}{2 m \si_0^2} + \frac{2 \si_0^2}{\hb} V_2 ,
\end{eqnarray}
where $ V_2 = d^2 V(x) / d x^2 $ is a constant for time-independent quadratic potentials.
One can thus rewrite this equation as a condition for the width initial value to have a soliton-like solution. 
Thus, Eq. (\ref{eq: soliton}) reveals that for free particles $\ga^{}_I$ must be negative in order to have a soliton-like solution. 
%

Concerning scaled Bohmian trajectories in the context of SL, let us start by introducing the polar form (\ref{eq: scaled_wf_polar}) of the scaled wave function 
into the scaled SL Eq. (\ref{eq: Scaled Sch-Lan-1}) and then decomposing into imaginary and real parts. One then
easily obtains  Eqs. (\ref{eq: tran_imag_part}) and (\ref{eq: tran_real_part}) with $\ti{\hbar}$ instead of $ \sqrt{\ep} \hbar$.
Then, from Eq. (\ref{eq: tran_imag_part}), the continuity equation is thus written as
\begin{eqnarray} \label{eq: bm1}
	\frac{\pa \ti{\rho}}{\pa t} + \frac{\pa}{\pa x} ( \ti{\rho} v_{\ep}) &=& 0 , 
\end{eqnarray}
where
\begin{eqnarray}
	\ti{\rho}(x, t) &=& R^2_{\ep}(x, t)
\end{eqnarray}
and
\begin{eqnarray} \label{eq: vel_field}
	v_{\ep}(x, t) &=& \frac{1}{m} \frac{\pa S_{\ep}(x, t)}{\pa x}
\end{eqnarray}
are the probability density and the corresponding velocity field, respectively.
From Eq. (\ref{eq: tran_real_part}) and the corresponding quantum potential defined by 
\begin{eqnarray}
Q_{\ep}(x, t) &=& - \frac{\hbar^2}{2m} \frac{1}{R_{\ep}} \frac{\pa^2 R_{\ep}}{\pa x^2}  ,
\end{eqnarray}
the Hamilton-Jacobi equation for the phase
\begin{eqnarray}
	- \frac{\pa S_{\ep}}{\pa t} &=&  
	\frac{1}{2m}  \left( \frac{\partial  S_{\ep}}{\partial x} \right)^2
	+ V + V_r + \gamma ( S_{\ep} - \langle S_{\ep} \rangle ) + \ti{Q} ,
\end{eqnarray}
is found  with  $\ti{Q} = \ep Q_{\ep}$.  The differential equation for the velocity field can be expressed as
\begin{eqnarray} \label{eq: bm2}
	\frac{dv_{\ep}}{dt}&=& \frac{\pa v_{\ep}}{\pa t} + v_{\ep} \frac{\pa v_{\ep}}{\pa x} = - \frac{1}{m} \frac{\pa}{\pa x} \left( V + V_r + \ti{Q}  \right) - \gamma v_{\ep}  ,
\end{eqnarray}
which is the classical equation of motion but with the additional term $\ti{Q}$ responsible 
for non-classical effects. One can solve Eqs. (\ref{eq: bm1}) and (\ref{eq: bm2}) 
by imposing a time-dependent Gaussian ansatz for the probability density of the form
\cite{NaMi-book-2017, ZaPlJD-AP-2015},
\begin{eqnarray} \label{eq: rho_ansatz1}
	\ti{\rho} (x, t) &=& \frac{1}{\sqrt{2\pi} ~ \ti{\sigma}(t)} \exp \left[ -\frac{(x-\ti q(t))^2}{2 \ti{\sigma}^2(t)} \right] ,
\end{eqnarray}
where $\ti q(t) = \int dx ~ x \, \ti{\rho}(x, t) $ is the time dependent expectation value of the 
position operator which follows the center of the Gaussian wave packet and $\ti{\sigma}(t)$ gives 
its width. 
Eq. (\ref{eq: rho_ansatz1}) satisfies  the continuity equation (\ref{eq: bm1}) for
\begin{eqnarray} \label{eq: Gauss-vel-field}
	v_{\ep}(x, t) &=& \frac{\dot{\ti{\sigma}}}{\ti{\sigma}} (x-\ti q(t)) + \dot{\ti q}(t) ,
\end{eqnarray}
from which scaled trajectories are generated and expressed as 
\begin{eqnarray} \label{eq: scaled-trajsield}
	x_{\ep}(x^{(0)}, t) &=& \ti q(t) + (x^{(0)} - x_0) \frac{\ti{\sigma}(t)}{\sigma_0} ,
\end{eqnarray}
with $x^{(0)}$ being the initial condition for the coordinate, $x_0$ the initial value for $x_t$ and $\sigma_0 = \ti{\sigma}(0)$. 
For $\epsilon = 1$, the standard Bohmian trajectories are recovered.
The same {\it dressing scheme} \cite{NaMi-book-2017} is again reached. Now, as usual, by replacing Eqs. (\ref{eq: rho_ansatz1}) 
and (\ref{eq: Gauss-vel-field}) into Eq. (\ref{eq: bm2}), and then Taylor expanding the interaction
potential around $\ti q(t)$ up to second order and using the condition for linear independence 
of different powers of $(x-\ti q(t))$, a classical Langevin equation for the position of the center 
of the wave packet and a second order differential equation in time for the width are then derived  
\begin{eqnarray}
	\ddot{\ti q}(t) + \gamma \dot{\ti q}(t) + \frac{1}{m} \left( F_r(t) + \frac{\pa V}{\pa x} \bigg\vert_{x=\ti q} \right) &=& 0 , \label{eq: xbar}  \\
	\ddot{ \ti{\sigma} } + \gamma \dot{ \ti{\sigma} } - \frac{\ti{\hbar}^2}{4 m^2 \ti{\sigma}^3} + 
	\frac{ \ti{\sigma} }{m} \frac{\pa^2 V}{\pa x^2} \bigg\vert_{x=\ti q} &=& 0 , \label{eq: delta1} 
\end{eqnarray}
where a linear form (\ref{eq: random_pot}) is again explicitely assumed for the random potential.
It is clear seen that $\ep$ and $\ga$ affect the width of the wave packet whereas, as expected from Ehrenfest's theorem, 
the  motion of its center is not altered by $\epsilon$ and, therefore, $\ti q(t)$ follows a 
classical trajectory.  Even more, due to the fact the scaled Planck's contant appears only in the differential
equation for the width, with $\ep$, the quantum character of its time evolution is gradually lost. 
Remember that the Gaussian ansatz is the exact solution of the transition wave equation for these cases where potentials of at most quadratic order are involved.

Moreover, when one neglects the random force term $F_r(t)$ and considers only dissipation with the second order interaction potential
written as
\begin{eqnarray} \label{eq: quad_taylor}
	V(x, t) & = & V_0(t) + V_1(t) x + \frac{1}{2} V_2(t) x^2 ,
\end{eqnarray}
then, from the previous two equations we have
\begin{eqnarray}
	\ddot{\ti q}(t) &=& - \gamma \dot{\ti q}(t) - \frac{V_1(t)}{m} - \frac{V_2(t)}{m} \ti q(t) . 
	\label{eq: xbar_quadratic} \\
	\ddot{ \ti{\sigma} } &=& - \gamma \dot{ \ti{\sigma} } + \frac{ \ti{\hbar}^2}{4 m^2 \ti{\sigma}^3} - 
	\frac{V_2(t)}{m} \ti{\sigma}  . \label{eq: delta_quadratic}  
\end{eqnarray}
%
Provided that $V_1$ and $V_2$ are time-independent, then the solution of  
Eq. (\ref{eq: xbar_quadratic}) is analytical and given by
\begin{eqnarray} \label{eq: xbar_quadratic_sol}
	x_t &=& - \frac{V_1}{V_2} + \left( x_0 + \frac{V_1}{V_2} \right) \left[ \cosh \Omega t + \frac{\gamma}{2} \frac{\sinh \Omega t}{\Omega} \right] e^{-\gamma t /2} + \dot{x}_0 ~ \frac{\sinh \Omega t}{\Omega}~e^{-\gamma t /2} , \nonumber \\
\end{eqnarray}
with the damped frequency
\begin{eqnarray} \label{eq: Omega}
	\Omega &=& \sqrt{-V_2/m + \gamma^2/4} .
\end{eqnarray}
On the contrary, the solution of  Eq. (\ref{eq: delta_quadratic}) is not found in an analytical 
way. However, for the non-dissipative or frictionless case $ \gamma = 0 $, provided that $V_2$ 
is independent of time, its solution is given by
\begin{eqnarray} \label{eq: sigma_nondis}
	\ti{\sigma}(t) &=& \sigma_0 \sqrt{ \cosh^2(\omega t) + \frac{\ti{\hbar}^2}{4 m^2 \omega^2 \sigma_0^4} \sinh^2(\omega t) } ,
\end{eqnarray}
for $\dot{\ti{\sigma}}(0) =0 $ and where $\omega = \sqrt{-V_2/m}$.
For the classical regime, $ \ep = 0$, and $V_2$ independent of time, the  solution of 
Eq. (\ref{eq: delta_quadratic}) is given by
\begin{eqnarray} \label{eq: delta_quadratic_cl}
	\sigma_{\cl}(t) &=& \sigma_0 \left( \cosh \Omega t + \frac{\gamma}{2} \frac{\sinh \Omega t}{\Omega} \right) e^{-\gamma t /2} + \dot{\sigma}_0 \frac{\sinh \Omega t}{\Omega} ~  
	e^{-\gamma t /2}  ,
\end{eqnarray}
which leads to 
\begin{eqnarray} \label{eq: delta_quadratic_cl_nondis}
	\sigma_{\cl}(t) &=& \sigma_0 ~ \cosh(\omega t) +  \dot{\sigma}_0 ~ \frac{ \sinh(\omega t) }{\omega}
\end{eqnarray}
in the non-dissipative case. In these equations, $ \dot{\sigma}_0 $ stands for the initial value of $\dot{\ti{\sigma}}(t)$. 

Now, from Eq. (\ref{eq: scaled-trajsield}), the  difference between two typical scaled trajectories
can be expressed as  
\begin{eqnarray} \label{eq: trajs_diff}
	x_{\ep}(x_1^{(0)}, t) - x_{\ep}(x_2^{(0)}, t) &=& (x_1^{(0)} - x_2^{(0)}) 
	\frac{\ti{\sigma}(t)}{\ti{\sigma_0}}~ .
\end{eqnarray}
Thus, trajectories diverge during the  time evolution revealing the non-crossing property of trajectories.
Notice that this property is even valid in the classical regime and can be used to provide a criterion  for tunnelling or, in general, 
any quantum effect.

\section{Results}\label{sec5}

\subsection{The Bohmian-Brownian motion} \label{subsec5-1}

The simplest open quantum system one can deal with  is the Brownian motion of a particle. When considering this motion in the Bohmian framework, this motion has been termed the Brownian-Bohmian motion \cite{NaMi-book-2017}.
For free propagation ($ V = 0 $), the stochastic classical trajectory (\ref{eq: xbar}) is written as
\begin{eqnarray} \label{eq: xt_free_formal_sol} 
	q(t) &=& q(0) + \dot{q}(0) \frac{1}{\ga} (1- e^{-\ga t} ) 
	+ \frac{1}{m \ga} \int_0^t dt'~ F_r(t') e^{-\ga ( t - t') } ,
\end{eqnarray}
where the random force has the properties of a Gaussian white noise;
\begin{eqnarray}
\la F_r(t) \ra &=& 0 , \\
\la F_r(0) F_r(t) \ra &=& 2 m \ga k_B T \det(t),
\end{eqnarray}
$ \del(t) $ being Dirac delta function, $k_B$ Boltzmann constant and $T$ temperature of the bath.
Then, from the standard properties  of the white noise and the Maxwell-Boltzmann (MB) distribution for the initial velocities, $ f_T(\dot{q}(0))  = \sqrt{ \frac{m}{2\pi k_B T} } \exp \left[ -\frac{m \dot{q}(0)^2}{2 k_B T} \right]$, one easily obtains the mean square displacement (MSD) to be
\begin{eqnarray} \label{eq: MSD_cl}
	\la\la ( q(t) - q(0))^2 \ra\ra & = & 2\frac{k_B T}{m \ga} \left( t - \frac{1-e^{-\ga t}}{\ga} \right)  ,
\end{eqnarray}
where the double averaging implies average over thermal fluctuations and  the initial velocities which are distributed according to 
the MB distribution function, $ \la \la \cdots \ra \ra = \int d\dot{q}(0) ~ f_T(\dot{q}(0)) ~ \la \cdots \ra $.
Eq. (\ref{eq: MSD_cl}) implies that in the diffusion regime $ t \gg \ga^{-1} $, the MSD is 
proportional to time with a constant given by $ 2 D $ where 
\begin{equation}
D =  \frac{k_B T}{m \ga} 
\end{equation}
is Einstein's law for the diffusion coefficient. A time-dependent 
diffusion coefficient $ D(t)$ can be defined as the ratio of MSD over $2t$ \cite{Mousavi-Salva-2019-1}.
From Eqs. (\ref{eq: BM-trajs}) and (\ref{eq: MSD_cl}), the 
MSD $ \la\la (x(x^{(0)}, t) - x^{(0)})^2 \ra \ra $ of stochastic Bohmian trajectories averaged over initial Bohmian positions $ x^{(0)} $ 
according to the Born rule gives the quantum diffusion coefficient \cite{Mousavi-Salva-2019-1}
\begin{equation} \label{eq: BM_D}
	D_{\qm}(t) = D_{\cl}(t) + \frac{1}{2 t} ( \si(t) - \si(0) )^2 , 
\end{equation}%
with the classical diffusion coefficient given by
\begin{equation}
	D_{\cl}(t) = \frac{k_B T}{m \ga} \left( 1 - \frac{1-e^{-\ga t}}{\ga t} \right)   .
\end{equation}
The quantum diffusion constant follows the same dressing scheme as the quantum trajectories in the Bohmian framework.
For the soliton-like solution Eq. (\ref{eq: soliton}), $D_{\qm}(t)$ is exactly the same as that of classical mechanics. 

%
\begin{figure} 
	\centering
	\includegraphics[width=10cm,angle=0]{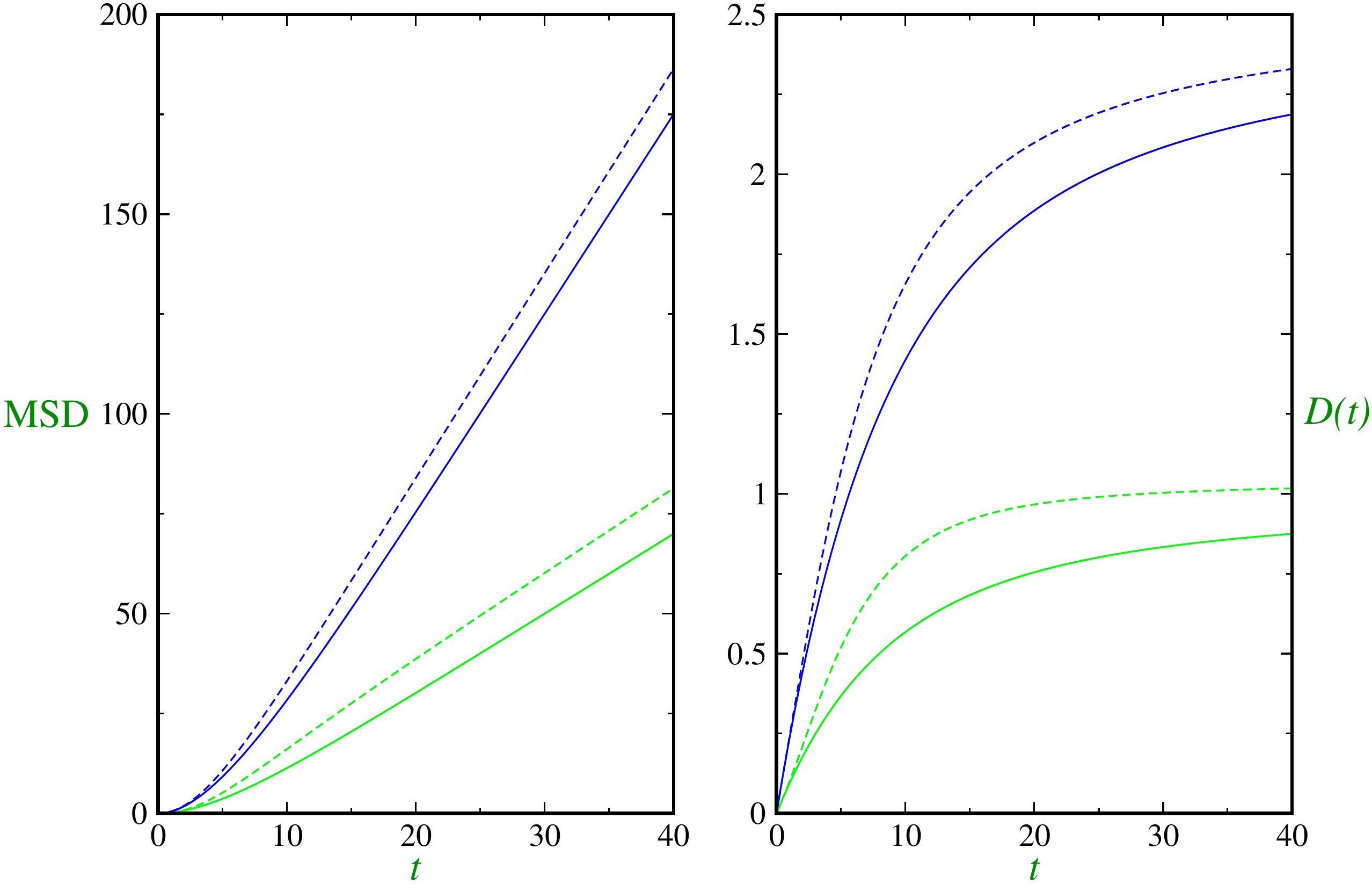}
	\caption{(Color online)
The time dependent MSD (left panel) and diffusion coefficient (right panel) are plotted for $\ga = 0.2$ for two different values of temperature; $ k_B T = 0.2 $ (green curves) and $ k_B T = 0.5 $ (blue curves). The dashed and solid curves are respectively for the quantum and classical regimes. Source: Adapted from Ref.  \cite{Mousavi-Salva-2019-1}}
	\label{fig: MSD&diff} 
\end{figure} 

In Fig. \ref{fig: MSD&diff}, we have plotted the time dependent MSD in (left panel) and the diffusion coefficient (right panel) for friction coefficient $\ga = 0.2$ and two different values of temperature; $ k_B T = 0.2 $ (green curves) and $ k_B T = 0.5 $ (blue curves).
As the right panel clearly shows each curve tends to an asymptotic value from 
which the corresponding diffusion coefficient can be extracted. From Einstein's relation, this coefficient increases with temperature.
Furthermore, quantum coefficients are always  greater than classical ones due to the width contribution of the wave packet according to 
Eq. (\ref{eq: BM_D}).

\subsection{Diffraction in time} \label{subsec5-2}

Sudden release from a totally absorbing shutter is of interest in the free dynamics leading to the so-called diffraction in time.
In this subsection, we want to analyze this diffraction in time in terms of dissipation within the scaled CK approach.

After sudden release from a confining potential, the physical system evolves freely. The dissipative propagator of the scaled CK equation (\ref{eq: Scaled CK}) is given by
\begin{eqnarray} \label{eq: prop_CK_scaled}
	G_f(x, t; x', 0) &=& \sqrt{ \frac{m}{ 2\pi i \ti{\hb} \tau(t)} } \exp \left[ \frac{i m  (x-x')^2 }{2 \ti{\hb} \tau(t)}\right]  ,
\end{eqnarray}
where
\begin{eqnarray} \label{eq:tau}
	\tau(t) &=&  \frac{ 1 - e^{-\ga t} }{ \ga }  .
\end{eqnarray}
As expected, this propagator reduces to that of the free particle for a non-dissipative environment. 

Suppose a beam of particles is trapped on the left side of an absorbing wall located at $x=0$. The initial unnormalized wave function is given by
\begin{eqnarray}
	\ti{\psi}_0(x) &=& e^{ i p x / \ti{\hb} } \theta(x)   ,
\end{eqnarray}
where $\theta(x)$ is the Heaviside function.
If the shutter is suddenly removed at $ t = 0 $, then the wavefunction is written as \cite{Mo-IJPR-2021}
\begin{eqnarray} \label{eq: psi_tilde}
	\ti{\psi}(x, t) &=& \frac{1}{\sqrt{2i}} \left( F(\ti{\xi}(x, t)) + \sqrt{ \frac{i}{2} } \right)
	\exp\left[ \frac{i}{\ti{\hb}} \left( px - \frac{p^2 \tau(t)}{2m} \right) \right]  ,
\end{eqnarray}
$ F(u) $ being the Fresnel integral defined as \cite{Gradshteyn-1980}
\begin{eqnarray} \label{eq: Fresnel}
	F(u) &=& \int_0^u dx~ e^{i\pi x^2 / 2}  ,
\end{eqnarray}
with
\begin{eqnarray}
	\ti{\xi}(x, t) &=& \sqrt{ \frac{m}{\pi \ti{\hb} \tau(t)} } \left[  \frac{p}{m}\tau(t) - x \right]  .
\end{eqnarray}
If we focus on the quantum regime where the transition parameter is $\ep=1$, Eq. (\ref{eq: psi_tilde}) gives the Moshinsky function. 
From this equation, one can write the probability density as
\begin{eqnarray} \label{eq: probden_shutter}
	\ti{\rho}(x, t) &=& \frac{1}{2} \left[ C(\ti{\xi}(x, t)) + \frac{1}{2} \right]^2
	+ \frac{1}{2} \left[  S(\ti{\xi}(x, t)) + \frac{1}{2} \right]^2  ,
\end{eqnarray}
where the functions $ C(u) $ and $ S(u) $ are the real and imaginary parts of the Fresnel function $ F(u) $, respectively. 
For a given position $ x = x_0 $, at time $ t = t_0 = - \frac{1}{\ga} \ln \left( 1 - \ga \frac{m x_0}{p} \right) $ where the argument of the Fresnel integral becomes zero, this density becomes $1/4$ for non-classical regimes.
In the long time limit $ \ga t \gg  1 $, $ \tau(t) $ becomes independent of time reaching the stationary value $1/\ga$ and the asymptotic value of the argument of the Fresnel integral is
\begin{eqnarray} \label{eq: Fres-argu-station}
	\ti{\xi}_{\infty}(x) & \approx & \sqrt{ \frac{m}{\pi \ti{\hb} \ga } } \left( \frac{p}{m\ga} - x \right)  .
\end{eqnarray}
Thus, the stationary value of the density is
\begin{eqnarray} \label{eq: PD-station}
	\ti{\rho}_{\infty}(x) &=& \frac{1}{2} \left( C( \ti{\xi}_{\infty}(x) ) + \frac{1}{2} \right)^2
	+ \frac{1}{2} \left( S( \ti{\xi}_{\infty}(x) ) + \frac{1}{2} \right)^2
\end{eqnarray}
this stationary value  being one for the non-dissipative case $ \ga = 0 $.

In the classical regime, $ \ep = 0 $, the probability density at a given location as a function of time is written as
\begin{eqnarray} \label{eq: PD-class}
	\rho_{\cl}(x, t) &=& 
	\begin{cases}
		0 & - \frac{1}{\ga} \ln \left( 1 - \ga \frac{m x}{p} \right) < t \\
		1 & t < - \frac{1}{\ga} \ln \left( 1 - \ga \frac{m x}{p} \right) .
	\end{cases}
\end{eqnarray}
This clearly shows that the classical density does not display the time oscillatory behavior as a function of time which is the hallmark of diffraction in time. 
\begin{figure} 
	\centering
	\includegraphics[width=10cm,angle=0]{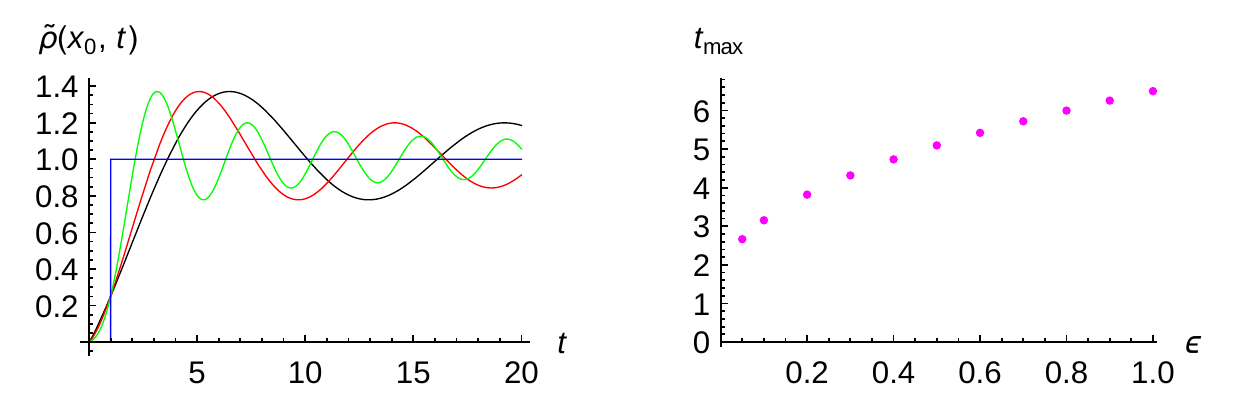}
	\caption{(Color online)
		Probability density for free friction, $\ga=0$, and initial position $x_0$ versus time in sudden removal of the shutter (left panel) and the location of the first maximum (right panel) for different regimes: quantum regime (black), transition regimes $\ep = 0.5$ (red) and 
		$\ep = 0.1$ (green); and the classical regime $\ep = 0$ (blue). For numerical calculations we have used $\hb=1$, $m=1$, 
		$p_0=1$ and $x_0=1$.  
	}
	\label{fig: DIT_CK} 
\end{figure} 

\begin{figure} 
	\centering
	\includegraphics[width=8cm,angle=0]{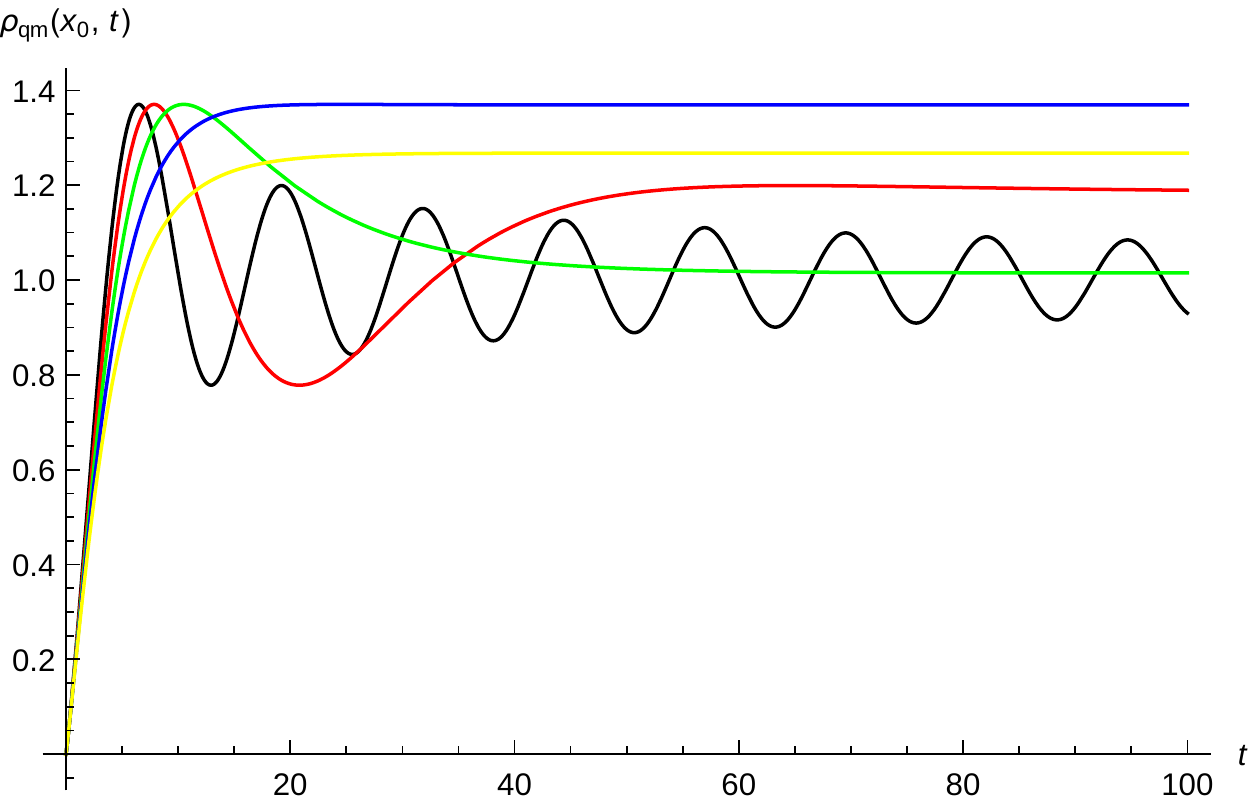}
	\caption{(Color online)
		Probability density in the quantum regime, $\ep=1$, in the given position $x_0$ versus time in sudden removal of the shutter (left panel) for different values of relaxation rate: $ \ga = 0 $ (black), $ \ga = 0.05 $ (red), $ \ga = 0.1 $ (green), $ \ga = 0.15 $ (blue) and $ \ga = 0.2 $ (yellow). For numerical calculations we have used $\hb=1$, $m=1$, $p_0=1$ and $x_0=1$.  
	}
	\label{fig: DIT_CK_qm} 
\end{figure}

In the left panel of figure \ref{fig: DIT_CK}, the characteristic oscillatory behavior, around the stationary value, is seen for several dynamical
regimes except for the classical one. In the quantum-classical transition, the number of oscillations increases and the pattern shifts to shorter times. 
In order to better observe these features, the location of the first maximum versus time for various regimes in the right panel of this figure is plotted. Values of the corresponding maxima and minima in various regimes are almost the same and thus the visibly, quantifying the
interference contrast, has almost the same value 0.5921 for all non-classical regimes.  
In Figure \ref{fig: DIT_CK_qm}, we have plotted the probability density in the quantum regime for different values of the relaxation rate or 
friction coefficient. As clearly seen,  the transient behaviors gradually disappear with $\ga$.

\subsection{Dissipative tunneling for a parabolic repeller under the presence of an electrical field} \label{subsec5-3}

Let us consider now  a charged particle being described  by a Gaussian wave packet (\ref{eq: rho_ansatz}). 
The interaction potential is given by
\begin{eqnarray} \label{eq: pot}
	V(x, t) &=& q E_0 \cos( \omega_0 t + \phi) ~ x - \frac{1}{2}  m \omega^2 x^2  ,
\end{eqnarray}
where $\omega$, $q$ and $m$ are the frequency, charge and mass of the harmonic
oscillator whereas $E_0$, $\omega_0$ and $\phi$ give the amplitude, frequency and phase of the applied field, respectively.
The corresponding equation of motion for the center of the Gaussian wave packet is given by
\begin{eqnarray} \label{eq: center_PROS}
	\ddot{x_t} + \gamma \dot{x_t} - \omega^2 x_t &=& - \frac{q E_0}{m}\cos( \omega_0 t + \phi), 
\end{eqnarray}
and the widths in the CK and Kostin  approaches are governed respectively by the following scaled dissipative Pinney equations
\begin{eqnarray} 
	\ddot{ \ti{\sigma} }  + \gamma \dot{ \ti{\sigma} } - \frac{ \ti{\hbar}^2}{4 m^2 \ti{\sigma}^3} e^{-2\gamma t} - \omega^2 \ti{\sigma} = 0, 
	\label{eq: CK_width_PROS} \\
	\ddot{ \ti{\sigma} }  + \gamma \dot{ \ti{\sigma} } - \frac{ \ti{\hbar}^2}{4 m^2 \ti{\sigma}^3} - \omega^2 \ti{\sigma} = 0 ,
	\label{eq: Kostin_width_PROS}
\end{eqnarray}
showing that the width of the Gaussian ansatz does not depend on the applied electrical field. The solution of 
Eq. (\ref{eq: center_PROS}) is 
\begin{eqnarray} \label{eq: xbar_quadratic_sol}
	x_t &=& \left[ x_0 \left( \cosh \Omega t + \frac{\gamma}{2} \frac{\sinh \Omega t}{\Omega}
	\right) + \dot{x}_0 ~ \frac{\sinh \Omega t}{\Omega} \right] e^{-\gamma t /2}
	\nonumber \\
	&+& e^{-\gamma t/2} \frac{q E_0 /m }{ \gamma^2 \omega_0^2 + ( \omega_0^2 + \omega^2  )^2 }
	\left[
	\left( \frac{\gamma^2}{2} + \omega_0^2 + \omega^2 \right) \frac{\sinh \Omega t}{\Omega} + \gamma \cosh \Omega t
	\right] \omega_0 \sin \phi
	\nonumber \\
	&+& e^{-\gamma t/2} \frac{q E_0 /m }{ \gamma^2 \omega_0^2 + ( \omega_0^2 + \omega^2  )^2 }
	\left[
	( \omega_0^2 - \omega^2 ) \frac{\gamma}{2} \frac{\sinh \Omega t}{\Omega} - ( \omega_0^2 + \omega^2 ) \cosh \Omega t
	\right] \cos \phi
	\nonumber \\
	&+&
	\frac{q E_0/m }{ \gamma^2 \omega_0^2 + ( \omega_0^2 + \omega^2  )^2 }
	\left[
	(\omega_0^2 + \omega^2) \cos (\omega_0 t + \phi)
	- \gamma \omega_0 \sin (\omega_0 t + \phi)
	\right],
\end{eqnarray}
where the expression in the last line is the particular solution of Eq. (\ref{eq: center_PROS}) with $\Omega$ given by
Eq. (\ref{eq: Omega}).
%
%

The time-dependent transmission probability for incidence from left to right  of the parabolic barrier is  known to 
be \cite{BaJa-JPA-1992,Pa-JPA-1990,Pa-JPA-1997}
\begin{eqnarray} \label{eq: tran_prob}
	T(t) &=& \frac{ B(t) }{ \int_{-\infty}^{x_m} dx~\rho(x, 0) }
\end{eqnarray}
with
\begin{eqnarray} \label{eq: B_coeff}
	B(t) &=& \int_0^t dt' ~ j(x_d, t') = \int_{x_d}^{\infty} dx~[\rho(x, t) - \rho(x, 0)]   ,
\end{eqnarray}
$x_m$ being the location of the barrier maximum, $x_m  = 0$. Here $x_d$ can be 
any point on the right side of the barrier. In the stationary regime, this probability reaches a constant value which is independent on 
the choice of $x_d$. The second equality in Eq. (\ref{eq: B_coeff}) results from integrating the continuity equation.
For $x_d = x_m = 0$, Eq. (\ref{eq: tran_prob}) reduces to 
\begin{eqnarray} \label{eq: tran_prob_gauss}
	T(t) &=& \frac{ {\text{erf}}( x_t / \sqrt{2} \ti{\sigma}(t) ) - {\text{erf}}( x_0 / \sqrt{2} \sigma_0 ) }{ {\text{erfc}}( x_0 / \sqrt{2} \sigma_0 ) }
\end{eqnarray}
for the wave packet (\ref{eq: rho_ansatz}). The transmission probability thus depends on the electrical
field through $x_t$  and  can be written in terms of the error function and its complementary function \cite{Gradshteyn-1980}.

\begin{figure}
	\centering
	\includegraphics[width=7cm,angle=-90]{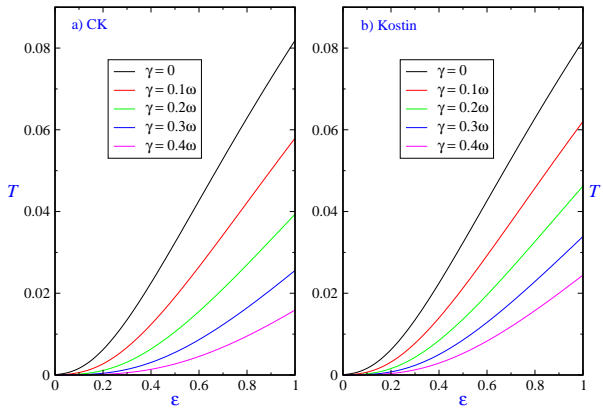}
	\caption{(Color online) 
		Transmission probability versus transition parameter $\epsilon$ for different values of 
		$\gamma$ within a) the CK and b) Kostin approaches. Source: Adapted from Ref. \cite{MoMi-AP-2018}}
	\label{fig: tranprob_ep}
\end{figure}

Numerical calculations have been carried out in atomic units ($m=1$, $\hbar=1$ and $q=-1$) and the initial Gaussian wave packet 
parameters are  chosen to be $\sigma_0=1$, $p_0 = 1$, $x_0 = -10$ and the frequency of  the parabolic repeller  $\omega = 0.2$. 
\begin{figure}
	\centering
	\includegraphics[width=8cm,angle=0]{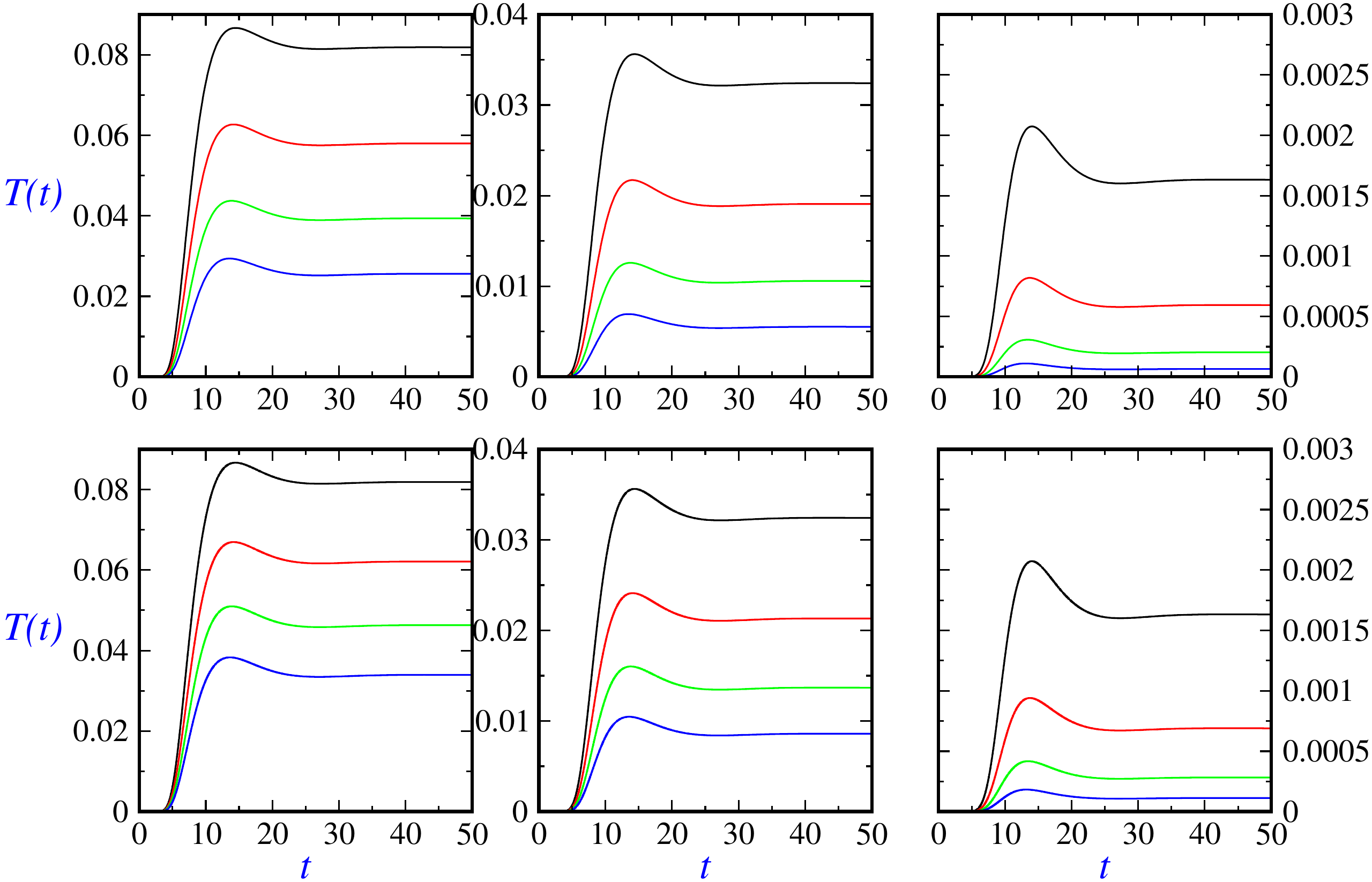}
	\caption{(Color online)
		Transmission probability as a function of time, $T(t)$, for seveal values of dissipation: 
		$\gamma = 0$ (black curve), $\gamma = 0.1 \, \omega$ (red curve), $\gamma = 0.2 \, \omega$ 
		(green curve) and $\gamma = 0.3 \, \omega$ (blue curve). This probability is plotted for several dynamical regimes, namely: 
		$\epsilon = 1$ (first column), $\epsilon = 0.5$ (second column) and $\epsilon = 0.1$ (third column)   
		in the  CK (first row) and Kostin (second row) approaches.
		Source: Adapted from Ref. \cite{MoMi-AP-2018}}
	\label{fig: tranprob_time}
\end{figure}
In Figure \ref{fig: tranprob_time}, transmission probabilities are displayed versus time for 
four different values of dissipation $\gamma = 0$ (black curve), 
$\gamma = 0.1 \, \omega$ (red curve), $\gamma = 0.2 \, \omega$ (green curve), and $\gamma = 0.3 \, \omega$ (blue curve) .
These probabilities are  plotted for different dynamical
regimes, namely: $\epsilon = 1$ (first column), $\epsilon = 0.5$ (second column) and $\epsilon = 0.1$ 
(third column)  in the  CK (first row) and Kostin (second row) approaches. Transmission probabilities 
decrease in both approaches and, with the transition parameter, the same is observed when passing 
from the quantum regime to a nearly classical regime ($\epsilon = 0.1$). 
Moreover, with this choice of parameters, 
the transmission probability becomes constant after $t \approx 30$ and displays a maximum at 
very short times.  This maximum corresponds to trajectories passing through the barrier but after a short period of time turn around. 
Kostin results always display a higher tunnelling transmission than the CK one as could be predicted from 
Eq. (\ref{eq: tran_prob_gauss}) since  $ \ti{\sigma}_{\text{CK}}(t) < \ti{\sigma}_{\text{SL}}(t) $.
This is also observed in Fig. \ref{fig: tranprob_ep} where these probabilities are plotted as a function of $\epsilon$
at four dissipative values for the two approaches. At around $t=150$, the asymptotic value of $T$ is already reached.
As expected, these probabilities decrease when $\epsilon$ tends to the classical limit and with $\gamma$ where the decoherence
process is playing more and more a major role.

\begin{figure}
	\centering
	\includegraphics[width=10cm,angle=0]{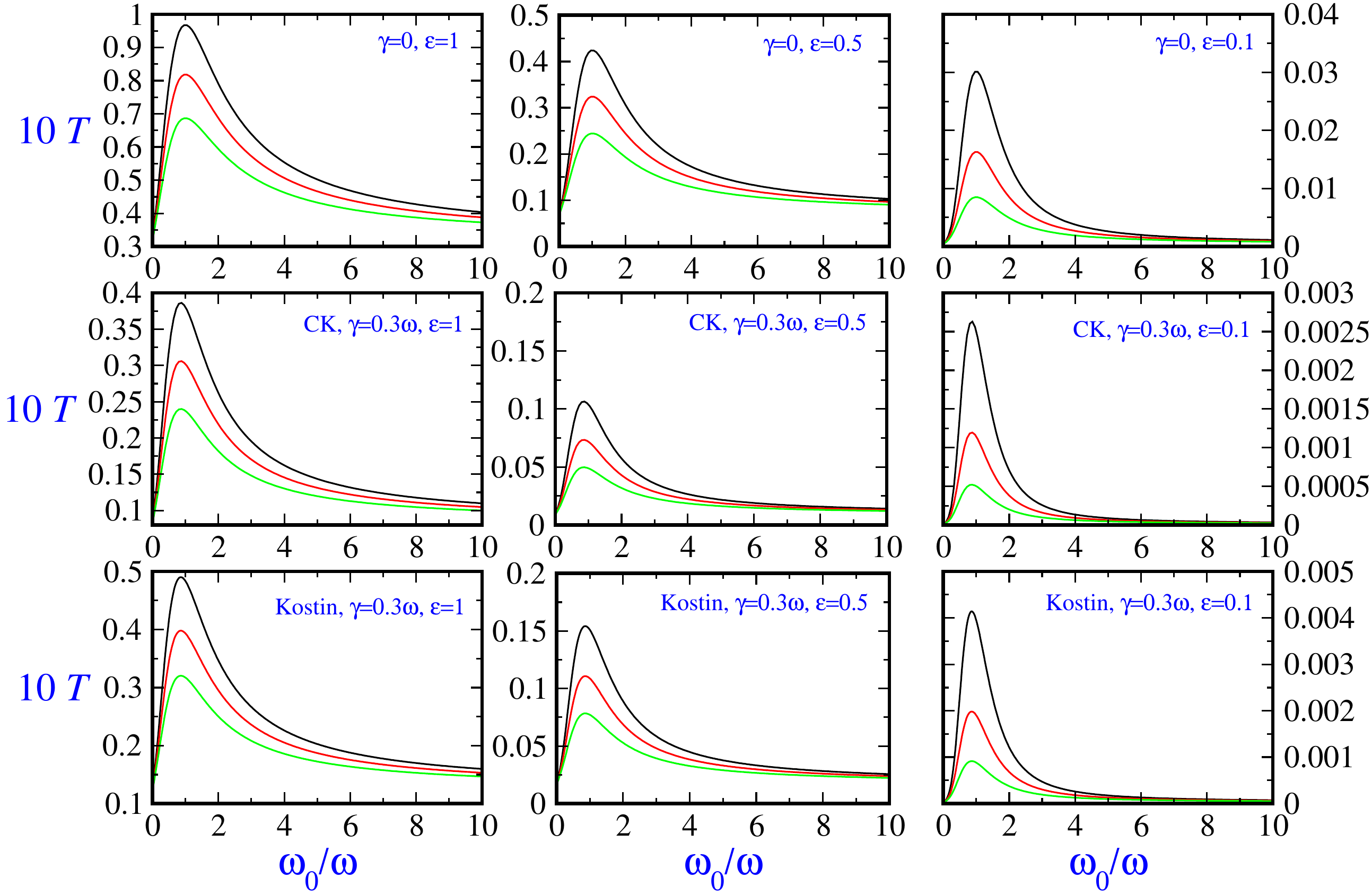}
	\caption{(Color online) 
		Transmission probability versus the frequency (in units of $\omega$) of the applied electrical field 
		for several amplitudes: $E_0 = 0.12$ (black curve), $E_0 = 0.1$ (red curve) and 
		$E_0 = 0.08$ (green curve) with $\phi=-\pi/2$. For comparison, in the first, second and third rows, the nondissipative
		motion, the CK  and Kostin approach (for $\gamma = 0.3 \omega$) are considered, repectively.
		The dynamical regimes studied are shown in the three columns with $\epsilon = 1$, 
		$\epsilon =0.5$ and $\epsilon =0.1$. Source: Adapted from Ref. \cite{MoMi-AP-2018}}
	\label{fig: tranprob_om0}
\end{figure}

\begin{figure}
	\centering
	\includegraphics[width=8cm,angle=-90]{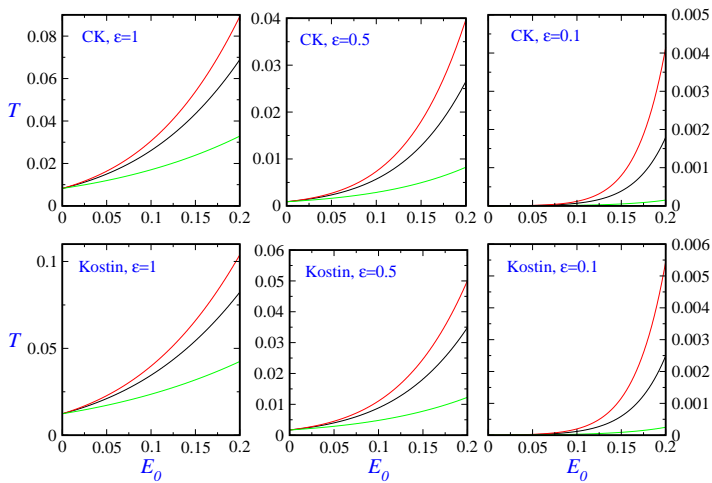}
	\caption{(Color online)
		Transmission probability versus the amplitude of the applied field for several values  
		of the frequency: $\omega_0 = 0.5 \omega$ (black curve), $\omega_0 = \omega_{0,\text{res}}
		\approx 0.86 \omega$ (red curve) and $\omega_0 = 3 \omega$ (green curve) 
		at different dynamical regimes in the two approaches, $\gamma = 0.3 \omega$ and $\phi=-\pi/2$. Source: Adapted from Ref. \cite{MoMi-AP-2018}}
	\label{fig: tranprob_E0}
\end{figure}

As can be seen, the variation of the tunnelling probability with the field parameters comes solely through $x_t$,  
the width $\ti{\sigma}(t)$ of the wave packet being independent on them. 
Figures \ref{fig: tranprob_om0} and \ref{fig: tranprob_E0}  display the dependence of transmission probability to the 
field parameters $\omega_0$ and $E_0$, respectively. 
Figure \ref{fig: tranprob_om0} shows the transmission probabilities versus the frequency 
(in units of $\omega$) of the applied electrical field for several amplitudes: $E_0 = 0.12$ (black curve), 
$E_0 = 0.1$ (red curve) and $E_0 = 0.08$ (green curve) with $\phi=-\pi/2$.  
With this initial phase, the field is a pure sine function of $(\omega_0 t)$. 
In the first, second and third rows, the nondissipative motion, the CK and Kostin approach
(for $\gamma = 0.3 \, \omega$) are plotted, respectively. 
The different dynamical regimes are also plotted in the three columns 
with $\epsilon = 1$, $\epsilon =0.5$ and $\epsilon =0.1$. A gradual decreasing of tunnelling is seen with
$\epsilon$, that is, when approaching the classical regime. In the absence of the field which this
occurs when $\omega_0 = 0$, the transmission probability is different from zero in the nonclassical
regime. The maximum observed is attributed to a resonant transmission and observed at any dynamical regime. As expected,
in the absence of friction, the  resonance takes place when $\omega_0 = \omega$. 
With friction, this  resonant mechanism is observed for $\omega_0 < \omega$. In our case, $ \omega_{0,\text{res}} \approx 0.86 \, \omega $.
Furthermore, the role of the field amplitude is just the opposite of  the friction, when increasing 
its value, the corresponding probabilities also increase in a nonlinear way. 
In the near classical regime $\epsilon = 0.1$, the tunnelling is very small. 
All of these features are also illustrated in Fig. \ref{fig: tranprob_E0} 
where these probabilities are plotted versus the amplitude of the applied field for 
different values  of the frequency: $\omega_0 = 0.5 \omega$ (black curve), 
$\omega_0 = \omega_{0,\text{res}}\approx 0.86 \, \omega$ (red curve) and 
$\omega_0 = 3 \omega$ (green curve) at different dynamical
regimes in the two approaches with $\gamma = 0.3 \, \omega$. The resonance or red curve gives the 
maximum value of the transmission probability.

\subsection{Early arrivals} \label{subsec5-4}

In time-dependent  barriers for  isolated systems, the scattering of wave packets  displays an effect known as  {\it early arrivals} 
\cite{HoMaMa-JPA-2012, superarrivals}. 
The transmission probability versus time increases when compared to the case of free wave packet propagation.
Thus, early arrivals refer to this early increase (relative to the free case) in the transmission probability. 
It is then of interest to study this effect for open quantum systems. To this end, we consider the corresponding transmission 
probabilities, in the context of the generalized Schr\"odinger equation (\ref{eq: gen_SCH2}), from a time-dependent parabolic barrier
\begin{eqnarray} \label{eq: para_rep}
	V(x, t) &=& - \frac{1}{2} m \om^2 e^{-g(t-t_B)^2} x^2  ,
\end{eqnarray}
which corresponds to the appearance of a parabolic repeller during a short time interval if a Gaussian time window is assumed. The 
parameters of the barrier are $ t_B $ and $g$ displaying the peak time and inverse width of the window, respectively. 
Here, $\om$ characterizes the strength of the barrier. 
If an initial wave packet is localized on the left side of the barrier, the transmission probability is then given by
\begin{eqnarray} \label{eq: tr_prob}
	P_{\tr}(t) &=& \int_{x_d}^{\infty} dx ~ \rho(x, t)
	= 
	\frac{1}{2} \mbox{erfc} \left[ \frac{x_d - q(t)}{\sqrt{2} \si(t)} \right]   ,
\end{eqnarray}
where $ x_d $ is the detector location and a Gaussian ansatz (\ref{eq: rho_ansatz}) is assumed. For the pure 
dissipative dynamics, one fixes $t_B$ from the condition $ q_f(t_B)  = 0 $ where $f$ denotes
the free case, The strength of the barrier is thus maximum when the center of the {\it free} Gaussian packet arrives at the top of the barrier. 
From the solution of Eq. (\ref{eq: xbar}) for the free dissipative dynamics, one has that
\begin{eqnarray}
	t_B &=& -\frac{1}{\ga^{}_R} \ln \left( 1 + \ga^{}_R \frac{q(0)}{\dot{q}(0)} \right)   .
\end{eqnarray}
Noting the negative sign of $q(0)$ and positive $ \dot{q}(0) $ and in order to have a positive time $ t_B $, one must impose the condition
\begin{eqnarray}
	\ga^{}_R \frac{\vert q(0) \vert}{\dot{q}(0)} &<& 1    .
\end{eqnarray}
\begin{figure} 
	\centering
	\includegraphics[width=10cm,angle=-90]{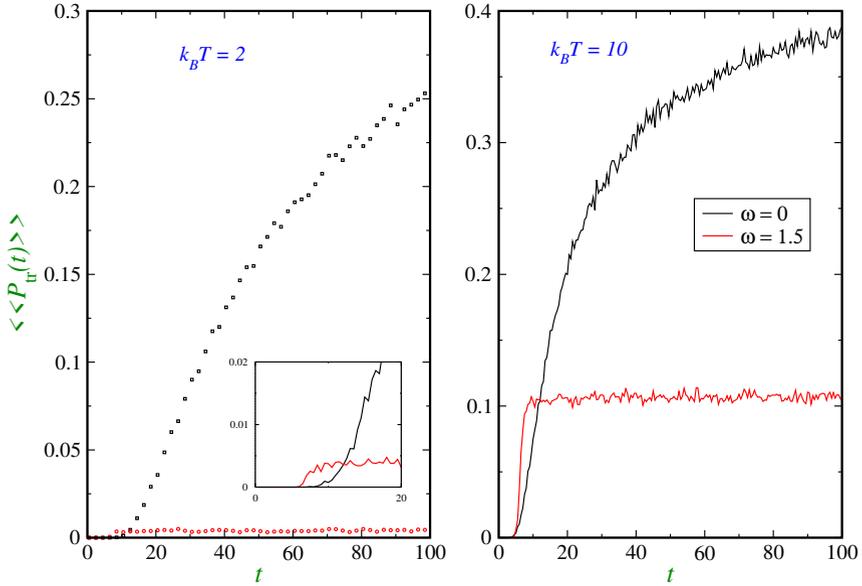}
	\caption{
		Time-dependent transmission probability under the presence of noise for $ k_B T = 2 $ (left panel) and $ k_B T = 10 $ (right panel) with
		$ \ga^{}_R = 0.1 $ and $ \ga_I = 0 $ for several values of barrier's strength: $\om=0$ (black curves) and $\om=1.5$ (red curves). The inset in the left panel 
		displays the early arrival effect. Source: Adapted from Ref. \cite{Mousavi-Salva-2019-1}
	}
	\label{fig: Trprob_noise} 
\end{figure}
In Figure \ref{fig: Trprob_noise}, the time-dependent transmission probability under the presence of thermal fluctuations or noise is plotted. 
The Langevin equation (\ref{eq: xbar}) is solved by using an algorithm proposed in Ref. \cite{VaCi-CPL-2006} with initial conditions 
$ q(t) \vert_{t=0} = q(0) $ and $ \dot{q}(t)\vert_{t=0} = \dot{q}(0) $ where $ \dot{q}(0) $ is chosen according to  
a Maxwell-Boltzmann distribution. Then, for the thermal transmission probability we have
\begin{eqnarray}
	\la \la P_{\tr}(t) \ra \ra &=& 
	\sum_{i=1}^{n_{\text{tra}}} P_{\tr, i}(t) =
	\frac{1}{2} \mbox{erfc} \left[ \frac{x_d - q_i(t)}{\sqrt{2} \si(t)} \right]  ,
\end{eqnarray}
where in the second equality we have used Eq. (\ref{eq: tr_prob}) and $q_i(t)$ refers to the $i-$th trajectory. The number of trajectories 
used to simulate a Maxwell-Boltzmann distribution for initial velocities is $ n_{\text{tra}} = 10,000 $.
Two different temperatures $ k_B T = 2 $ (left panel) and $ k_B T = 10 $ (right panel) are used for $ \ga^{}_R = 0.1 $ and $ \ga^{}_I = 0 $. 
In each panel, two different values of barrier's strength $\om=0$ (black curves) and $\om=1.5$ (red curves) are showed.
Here, we have set $ t_B $ in Eq. (\ref{eq: para_rep}) as $ t_B = 3 t_b $ where $ t_b = 2m\si_0^2/\hb $ is the time-scale appearing in the 
relation of freely propagating Gaussian wave packet in a non-dissipative medium.
Therefore, the effect of early arrivals is also seen in this stochastic dynamics.
One clearly sees that temperature also enhances transmission probability. 
%

\section{Conclusions} \label{sec6}

The main purpose of this work is to {\it review} the dynamics of open quantum systems by means of two approaches 
which are less used and applied in this context than the master equation for the reduced density matrix namely,
effective Hamiltonians and nonlinear Schr\"odinger equations. The decoherence process is described in terms of trajectories
in the configuration space inspired by Bohmian mechanics, leading to a more intuitive way of understanding the 
underlying physics. This theoretical analysis is carried out by calculating dissipative/stochastic
Bohmian and scaled trajectories in order to better understand the quantum-to-classical transition. This transition
is monitored by the so-called transition parameter $\epsilon$ which goes from $\epsilon = 0$ (classical regime) to $
\epsilon = 1$ (quantum regime). As mentioned in the Introduction, this gradual transition can be seen as an
alternative way to the WKB approximation (valid for conservative systems) but in the context of open quantum dynamics. 
An important feature of this
analysis is the theoretical observation that using trajectories, the so-called dressing scheme is kept for all types of 
trajectories, similar to the soliton description of nonlinear quantum mechanics. Furthermore, we have reached 
for all cases presented here different generalized continuity, Hamilton-Jacobi and Pinney equations.
In the Introduction we mentioned the two sources of decoherence of different nature presented here. The transition parameter
facilitate us to analyze continuously dynamical regimes between the quantum and classical ones but keeping always the open character
of them. Both sources of decoherence work in the same direction.

In order to illustrate the theory behind the two approaches used, several well-known problems have been considered.
First,  the Brownian-Bohmian motion which is the paradigm of open quantum systems (or quantum
Brownian motion). In this stochastic dynamics, a quantum diffusion coefficient is obtained in terms of the standard classical one. 
Einstein's law for diffusion is reproduced. The second problem is diffraction 
in time which is an effect occurring  by sudden release from a totally absorbing shutter in a free dynamics. The hallmark of this effect
is still robust for dissipative dynamics. The third problem is
the study of dissipative tunnelling by an inverted parabolic barrier 
showing how the decoherence process is  gradually established for the different dynamical regimes and friction values.
The role played by electrical field parameters (in particular, frequency and amplitude) shows that
the resonance process observed is valid for any dynamical regime, except the pure classical one.
An interesting point here is that when comparing the Kostin and CK approaches, the 
tunnelling probabilities are different as also observed by Tokieda and Hagino 
\cite{ToHa-PRC-2017}. This discrepancy comes from the differential equation 
governing the Gaussian width used where both approaches differ. 
The Kostin approach seems to be more convenient than the CK one for at least two reasons. First,
when an interaction with an environment (bath, measuring apparatus, etc) is present, 
linear quantum mechanics seems to be no longer applicable and nonlinear differential equations have to 
be implemented for a proper description of the corresponding open quantum dynamics. And second,
the Kostin equation comes from the quantum Langevin equation which is also issued from
a Caldeira-Leggett Hamiltonian formalism, whereas the CK Hamiltonian comes from 
a phenomenological or effective approach.  Finally, the fourth problem is the well-known early arrivals effect
for time dependent barriers.
From stochastic trajectories, we have shown again that this effect is still robust under the presence of
friction and temperature.

The success of the different trajectories applied to these simple problems should be seen as an alternative
way to the master equation and path integral  formalisms as well as a stimulus to
extend this type of analysis to more general problems in order to see the validity and range of application 
of the different nonlinear Schr\"odinger equations. In particular, the solution of the SL nonlinear equation (or similar equations
analyzed here or in the literature) could be explored following, for example,  the well-known stochastic wave function method
\cite{Percival1998,BaPe-book-2002}.

\vspace{2cm}
\noindent
{\bf Acknowledgements}
SVM acknowledges support from the University of Qom. SMA would like to thank Prof. A. B. Nassar for their short but fruitful collaboration on this subject. SMA acknowledges support from 
the Ministerio de Econom\'ia y Competitividad (Spain) under the Project 
PID2021-125735NB-I00 and Fundaci\'on Humanismo y Ciencia.




\end{document}